\documentclass[twocolumn,showpacs,eqsecnum,aps,nofootinbib,preprintnumbers,floatfix]{revtex4}

\usepackage{citesort,graphicx}

\def  \btosll      {$b \to s \ell^+ \ell^-$ }
\def  \btoxsll     {$B \to X_s \ell^+ \ell^-$ }
\def  \btokstartt  {$B \to K^* \tau^+ \tau^-$ }
\def  \btoxstt     {$B \to X_s \tau^+ \tau^-$ }

\def  \bcen        {\begin{center}}
\def  \ecen        {\end{center}} 
\def  \beq         {\begin{equation}}
\def  \eeq         {\end{equation}}
\def  \beqa        {\begin{eqnarray}}
\def  \eeqa        {\end{eqnarray}}
\def  \bfig        {\begin{figure}}
\def  \efig        {\end{figure}} 

\def  \vtbvts      {V_{tb} V_{ts}^* }

\def  \sh          {\hat{s} }
\def  \mlh         {\hat{m}_\ell }

\def  \cseveff     {C_7^{eff} }
\def  \cneff       {C_9^{eff} }
\def  \cten        {C_{10} }
\def  \cqone       {C_{Q_1} }
\def  \cqtwo       {C_{Q_2} }

\def  \ev#1        {{\bf e}_#1}
\def  \wv#1        {{\bf w}_#1}
\def  \pmv         {{\bf p_-}}
\def  \ppv         {{\bf p_+}}
\def  \psv         {{\bf p_s}}  

\def  \etal        {{\it et. al.} }

\def \prd#1#2#3       {Phys. \ Rev. {\bf D #1}, {#2} (#3)}
\def \prl#1#2#3       {Phys. \ Rev. \ Lett. {\bf #1}, {#2} (#3)}
\def \nuclphysb#1#2#3 {Nucl. \ Phys. {\bf B #1}, {#2} (#3)}
\def \plb#1#2#3       {Phys. \ Lett. {\bf B #1}, {#2} (#3)}
\def \physrep#1#2#3   {Phys. \ Rep {\bf #1}, {#2} (#3)}
\def \zphysc#1#2#3    {Z. \ Phys. {\bf C #1}, {#2} (#3)}

\begin{document}

\preprint{
hep-ph/0305242
}

\title{\boldmath Lepton Polarization asymmetries in \btoxstt in MSSM }

\author{Naveen Gaur}
  \email{naveen@physics.du.ac.in }
  
  \affiliation{Department of Physics \& Astrophysics,
University of Delhi, 
Delhi - 110 007, India}

\date{\today}



\begin{abstract}
Semi-leptonic and leptonic decays of B-mesons are important probes for
testing SM and theories beyond it because of their relative
cleanliness and far less theoretical uncertainties. In semi-leptonic
decays based on quark level transition \btosll apart from branching
ratio one can study many other (possible) observables associated with
final state leptons like, lepton pair Forward Backward asymmetry,
lepton polarization asymmetries etc. But as proposed recently if we
can tag the B-meson than one can measure the polarization asymmetries
of both the leptons. Here we will study the polarization asymmetries
of both the final state leptons in SM and Minimal Supersymmetric
Extension (MSSM) to it 
\end{abstract}

\maketitle


\section{\label{section:1}Introduction}
The rare B-meson decays induced by Flavor Changing Neutral Current
(FCNC) $b \to s(d)$ transitions arises only at loop levels in Standard
Model (SM). Because these transitions occur at loop level, hence
provides useful tests of the detailed structure of the theory at the
levels where GIM (Glashow-Iliopolus-Maini) cancellations becomes very
important. Also in most of the extensions of SM, loop graphs with new
particles (most of the extensions of SM predict existence of some new
particles like SUSY predicts whole set of SUSY particles) can
contribute to the same order (in case of pure dileptonic decay of
$B_s$ meson SUSY can give a contribution which is many orders greater
than SM contribution \cite{Choudhury:1999ze,Baek:2002wm}). In
particular the process $B \to X_s \gamma$ and \btoxsll are
experimentally very clean and can possibly be more sensitive to any
new Physics beyond the SM. The new physics effects in the rare decays
can come in two ways, 
one via the modifications of the existing (in SM) Wilson coefficients 
\cite{Cho:1996we} 
and other via the introduction of some new operators (accompanied by
new coefficients) \cite{Xiong:2001up,Baek:2002wm}. The inclusive decay
mode (\btoxsll) can be more sensitive than the radiative decay mode
($B \to X_s \gamma$) for  testing any new physics model because in
inclusive decay mode many more kinematical distributions like, lepton
pair forward backward (FB) asymmetry, leptons polarization asymmetry
etc. can be measured.  

\par Various kinematical distributions of the inclusive mode have been
studied in many earlier works
\cite{Xiong:2001up,RaiChoudhury:1999qb}. As also been proved in many
of the works that lepton polarization asymmetry of final state leptons
can give us useful information to fit parameters of SM and constraint
new physics models \cite{RaiChoudhury:1999qb,Kruger:1996cv}. But
recently it has been noted down, in case of inclusive decays by
Bensalem et.al, \cite{Bensalem:2002ni}
that one can in principle observe many more observables, like the
double polarization asymmetries (polarization asymmetries when both
the leptons are polarized), which would be useful in further testing
of the SM and probing physics beyond it. 
 They inferred that even if we won't be able
to tag b but can observe the final polarization state of both
the leptons than also we can have more distributions than what we have
if we would only be measuring the single lepton polarization
asymmetries. But if we can
also have {\sl b-tagging} along with the measurement of the
polarization of both the final state leptons, then many more
kinematical distributions 
would be available to us, which could be useful in testing the
structure of effective Hamiltonian and hence the physics underlying
it. In our earlier work \cite{Choudhury:2003a} we tried to estimate as
to what would be the form of various double polarization distributions
within SM for the exclusive mode \btokstartt and how these
distributions gets modified 
by switching on the SUSY effects. This work is the extension of the
work of Bensalem et.al., \cite{Bensalem:2002ni}. In their work they
restricted themselves to SM operators. In this work we will extend the
SM operator basis and will try to observe the aftereffects of this on
the various double polarization asymmetries.  

\par The quark level transition we would be interested in the present
work is \btosll. But as has been very well emphasized in literature
that if we consider the Supersymmetric (SUSY) extension of the
SM, than we have to extend the SM list of operators to include the
operators arising from the exchange Neutral Higgs Bosons (NHBs)
\cite{Choudhury:1999ze,Baek:2002wm,Xiong:2001up,RaiChoudhury:1999qb,Skiba:1993mg}.
In case of pure-dileptonic decays of B-meson ($B_s \to \ell^+   
\ell^-$) the NHBs can change the SM predictions by many orders of
magnitude \cite{Choudhury:1999ze,Baek:2002wm,Skiba:1993mg}. As the
couplings of the NHBs to leptons is proportional to the lepton mass,
hence these effects becomes more important if $\ell = \mu, \tau$. The
SUSY extension of the SM predicts existence of two new operators,
which were not present within SM. These new operators are responsible
for the orders enhancement of the branching ratio of pure dileptonic
decay mode of B-meson. In our analysis we will be going to consider
these operators and will try to estimate the dependence of the various
double polarization asymmetries on MSSM parameters. 

\par The paper is organized as follows : In section \ref{section:2}
we will present the effective Hamiltonian we are considering. We will
then write the matrix element for the quark level process \btosll. In
section \ref{section:3} we will give the definition and the
analytical results of the double polarization asymmetries. Finally, we
will conclude in section \ref{section:4} by discussing our numerical
results. 


\section{\label{section:2} Effective Hamiltonian}

\begin{figure}
\vskip 0.3cm
\includegraphics[width=3in]{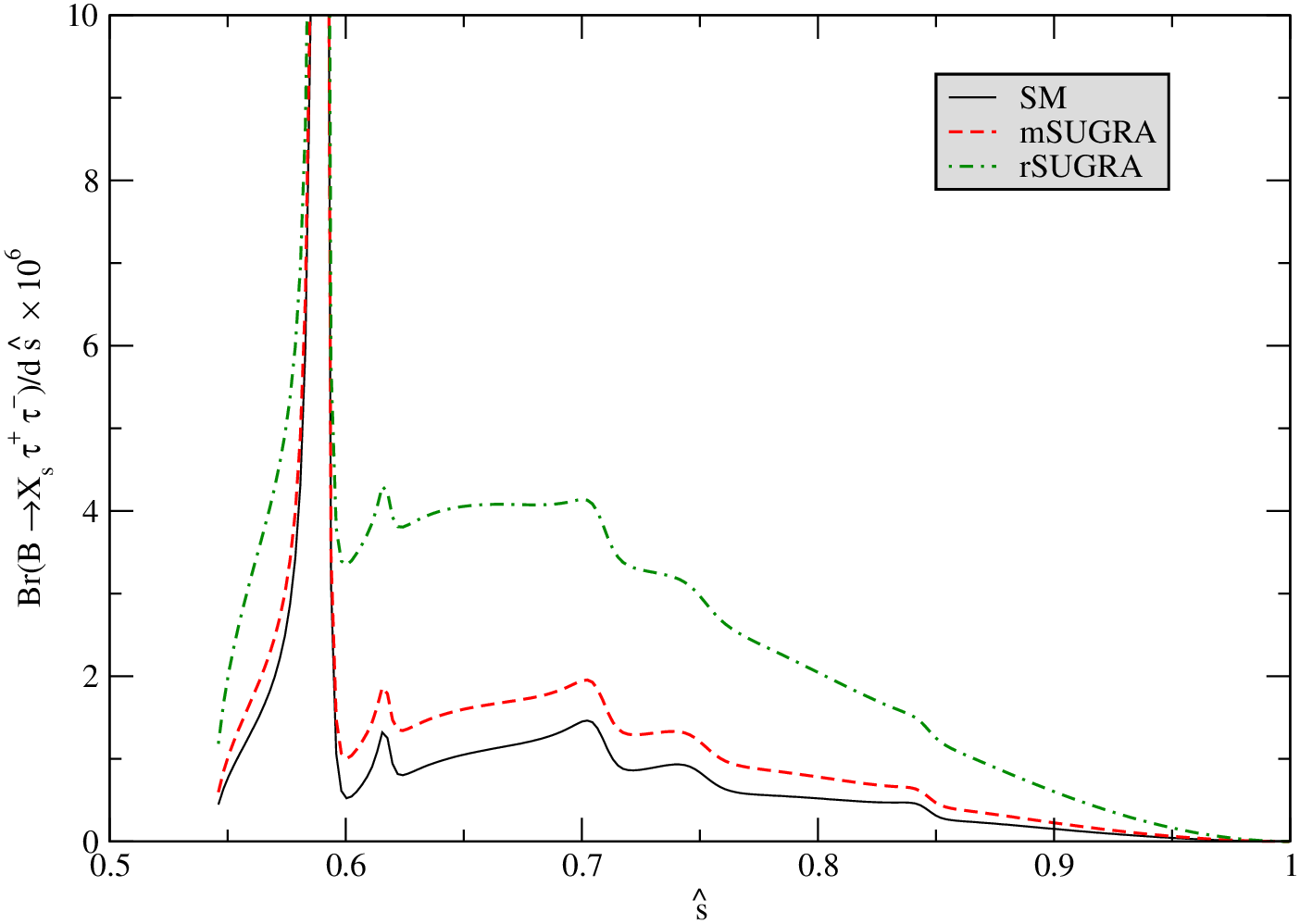}
\vskip -0.2cm
\caption{\label{fig:dr_s}Branching ratio of \btoxstt with scaled
invariant mass of dileptons. Parameters of mSUGRA model are : $m =
400$GeV, $M = 500$GeV, $A = 0$, $tan\beta = 40$ and $sgn(\mu)$ is
positive. The additional parameter for rSUGRA is $m_A = 270$GeV }
\end{figure}

We follow the convention followed in
\cite{Xiong:2001up,RaiChoudhury:1999qb} to write down the matrix
element and invariant mass spectrum. The inclusive decay mode 
(\btoxsll) is modeled by partonic process $b(p_b) \to s(p_s) +
\ell^+(p_+) + \ell^-(p_-)$. This sort of modeling is actually the
leading order calculation in $1/m_b$ expansion. By integrating out the
heavy degrees of freedom from the full theory (MSSM here) we can get
the effective Hamiltonian describing the semi-leptonic decay \btosll
\cite{Choudhury:1999ze,Baek:2002wm,Xiong:2001up,RaiChoudhury:1999qb} :
\beq
{\cal H}_{eff} = \frac{4 G_F}{\sqrt{2}} \vtbvts 
\left(
  \sum_{i=1}^{10} C_i O_i + \sum_{i=1}^{10} C_{Q_i} Q_i
\right)
\label{sec2:eq:1}
\eeq
where $O_i$ are current-current (i=1,2), penguin (i = 1,\dots,6),
magnetic penguin (i=7,8) and semi-leptonic (i = 9,10) operators and
$C_i$ are the corresponding Wilsons. They have been given in
\cite{Cho:1996we,Grinstein:1989me}. The additional operators $Q_i$ (i
= 1,\dots,10) and their Wilson coefficients ($C_{Q_i}$) which arises
due to MSSM diagrams are given in
\cite{Choudhury:1999ze,Xiong:2001up}.   

\par Neglecting the mass of s-quark, the effective Hamiltonian gives
the matrix element :
\beqa
{\cal M} = &&\frac{\alpha G_F}{\sqrt{2} \pi} \vtbvts
\Bigg\{
 - 2 \cseveff \frac{m_b}{q^2} (\bar{s} i \sigma_{\mu \nu} q^\nu P_R b)
( \bar{\ell} \gamma^\mu \ell )  
    \nonumber \\
&& 
 + \cneff (\bar{s} \gamma_\mu P_L b) (\bar{\ell} \gamma^\mu \ell)
 + \cten (\bar{s} \gamma_\mu P_L b) (\bar{\ell} \gamma^\mu \gamma_5
  \ell)  
  \nonumber \\
&&
  + \cqone (\bar{s} P_R b) (\bar{\ell} \ell)
 + \cqtwo (\bar{s} P_R b) (\bar{\ell} \gamma_5 \ell) 
\Bigg\}
\label{sec2:eq:2}
\eeqa
where q is the momentum transfer to the lepton pair given as $q = p_+ +
p_-$, where $p_-$ and $p_+$ are the momentas of $\ell^-$ and $\ell^+$
respectively. $\vtbvts$ are the CKM factors and $P_{L,R} = (1 \mp
\gamma_5)/2$ . 

\par The matrix element ${\cal M}(b \to s \ell^- \ell^+)$ is the free
quark decay amplitude but it has some long distance effects also which
comes due to the four-quark operators $\langle \ell^+ \ell^- s| O_i |b
\rangle$ where $i = 1,\dots,6$. These effects are absorbed into the
redefinition of short distance Wilson coefficients according to the
prescription given in many earlier works
\cite{Kruger:1996cv,Long-Distance}. The long distance effects are
because of the the $c \bar{c}$ resonance contributions. These
resonances can be taken into account by using Breit-Wigner ansatz
\cite{Kruger:1996cv,Long-Distance} by which we add a term to $\cneff$.
The resonance contribution to $\cneff$ is :
\beq
C_9^{res} \propto \kappa \sum_{V = \psi} 
         \frac{\hat{m}_V Br(V \to \ell^- \ell^+) 
       \hat{\Gamma}^V_{total}}{\sh - \hat{m}_V^2 + i \hat{m}_V
        \hat{\Gamma}^V_{total}}
\label{sec2:eq:3}
\eeq
where the symbols are explained in Kruger \& Sehgal
\cite{Kruger:1996cv}. The phenomenological factor $\kappa$ has to be
introduced to reproduce the correct branching ratio ${\cal B}(B \to
J/\Psi X_s) \to X_s \ell^+ \ell^-)$. We will be going to take this
factor to be 2.3 for our numerical analysis. 

\begin{figure}
\vskip 0.5cm
\includegraphics[width=3in]{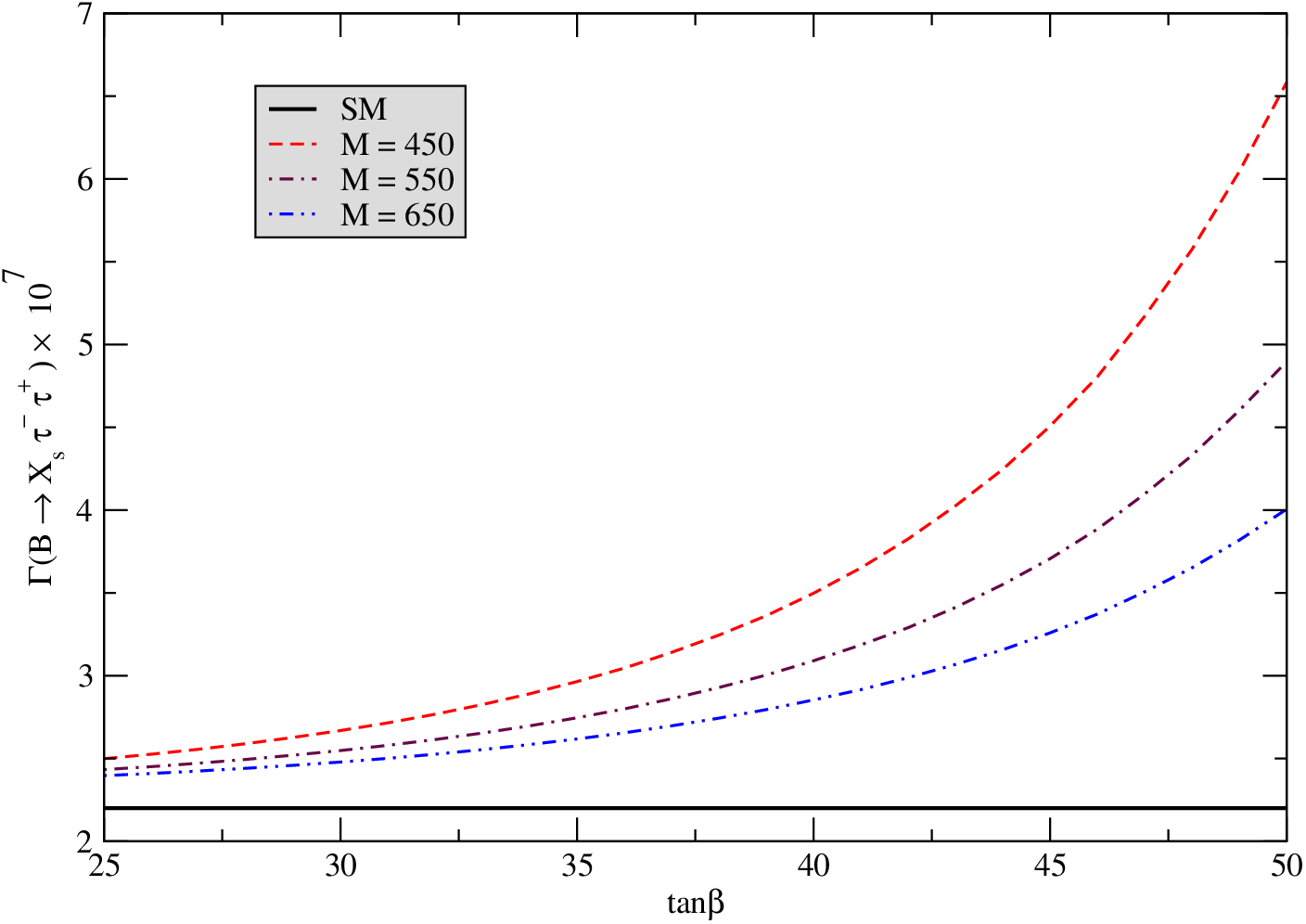}
\caption{\label{fig:dr_tb}Total decay rate with $tan\beta$ in mSUGRA
model. Other parameters $m = 400$GeV, $A = 0$}
\end{figure}

From the expression of the matrix element given in
eqn.(\ref{sec2:eq:2}) we can get the dilepton invariant mass
distribution as : 
\beq
\frac{d \Gamma}{d \sh} = \frac{G_F m_b^5}{192 \pi^3} \frac{\alpha^2}{4
\pi^2} |V_{tb} V_{ts}^*|^2 (1 - \sh)^2 \sqrt{1 - \frac{4 \mlh^2}{\sh}}
\bigtriangleup
\label{sec2:eq:8}
\eeq
with
\beqa
\bigtriangleup &=&
4  \frac{(2 + \sh)}{\sh} \left(1 + \frac{2 \mlh^2}{\sh}\right)
 |\cseveff|^2 + (1 + 2 \sh) 
         \nonumber  \\
  && \left(1 + \frac{2 \mlh^2}{\sh}\right) |\cneff|^2 
+ (1 - 8 \mlh^2 + 2 \sh + \frac{2 \mlh^2}{\sh}) \nonumber \\
&& \times |\cten|^2 + {3 \over 2} (-4 \mlh^2  + \sh) |\cqone|^2
+ {3 \over 2} \sh |\cqtwo|^2   \nonumber \\
&& + 12 (1 + \frac{2 \mlh^2}{\sh})  
 Re(C_9^{eff *} \cseveff)  \nonumber \\
&& + 6 \mlh  Re(\cten ^* \cqtwo)
\label{sec2:eq:9}
\eeqa

\begin{figure}
\vskip 0.5cm
\includegraphics[width=3 in]{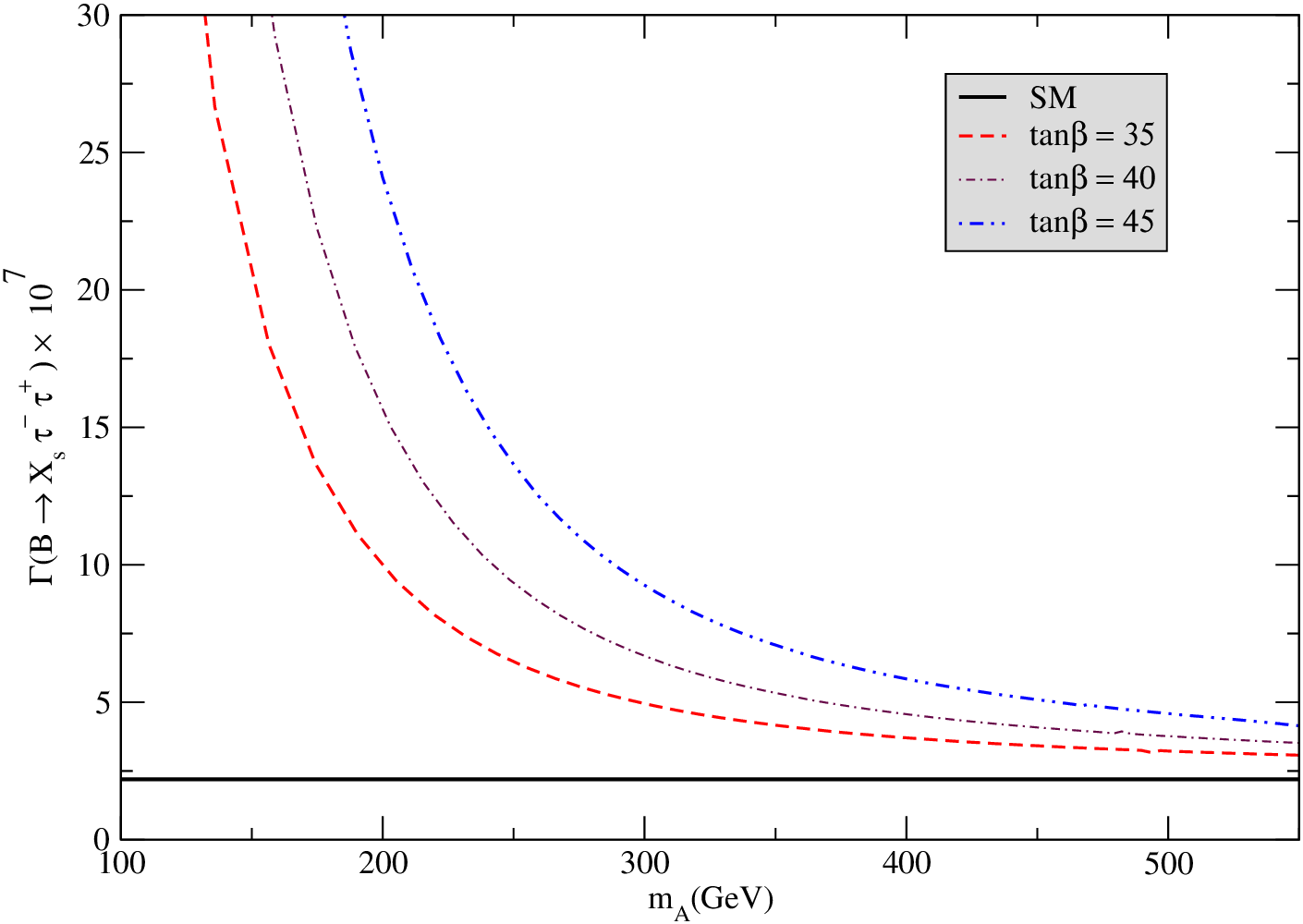}
\caption{\label{fig:dr_ma}Total decay rate with $m_A$GeV in rSUGRA
model. Other parameters $m = 400$GeV, $M = 500$ GeV, $A = 0$}
\end{figure}

Now ready with the expression of the invariant mass spectrum
(including the scalar exchange effects) we will analyze various double
polarization asymmetries in next section. As has been emphasized in
many texts that the $\tau$-polarization asymmetries in inclusive decay
\btoxstt can be a very useful probe of the structure of effective
Hamiltonian and hence the underlying theory. So for our analysis we
will consider the inclusive decay channel with $\tau$ leptons in final
state.


\section{\label{section:3} Lepton polarization asymmetries}

\begin{figure}
\vskip .6cm
\includegraphics[width=3in]{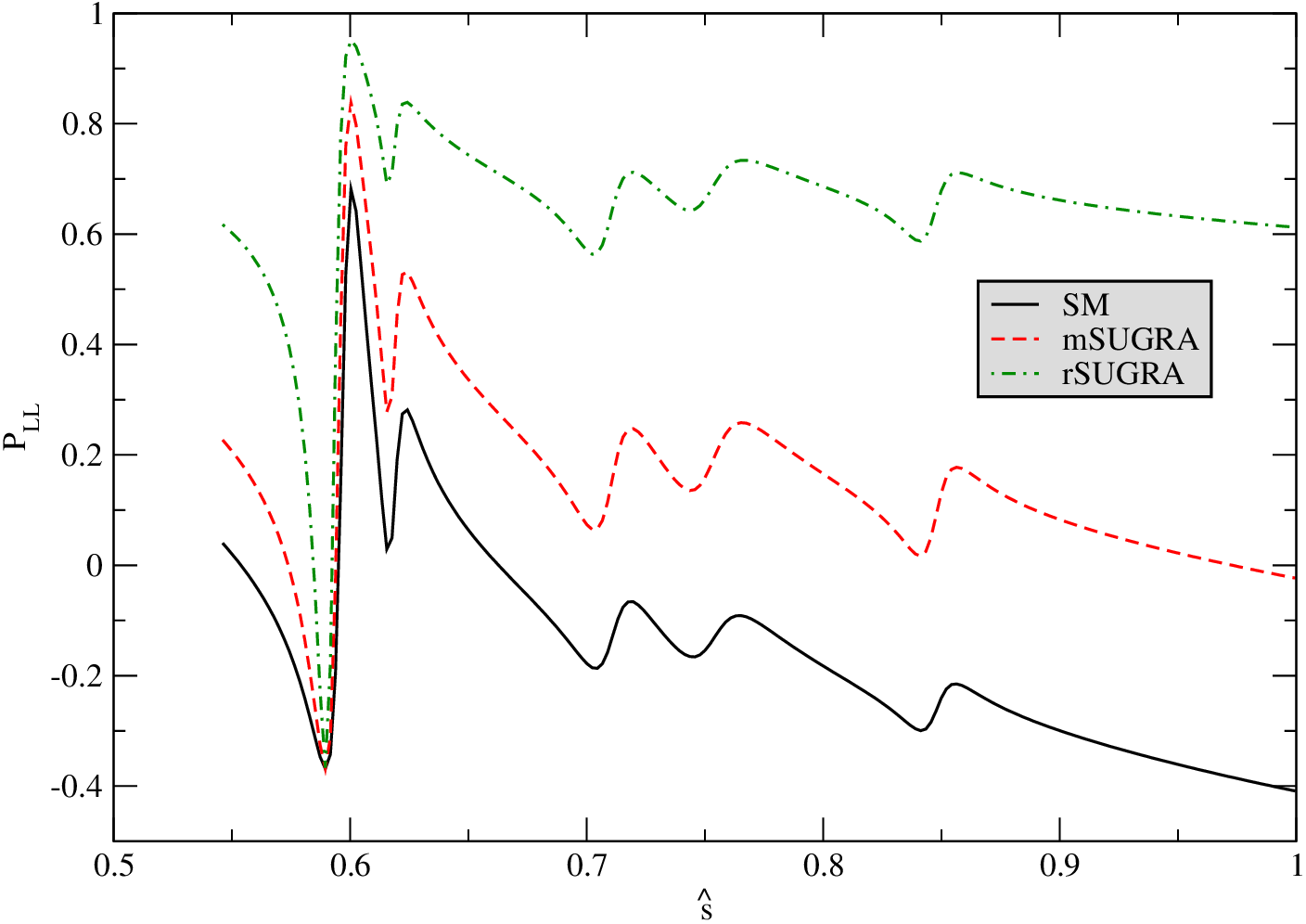}
\caption{\label{fig:pll_s}${\cal P}_{LL}$ with $\sh$ with all the
parameters same as in Fig. \ref{fig:dr_s} }
\end{figure}
\begin{figure}
\vskip .6cm
\includegraphics[width=3in]{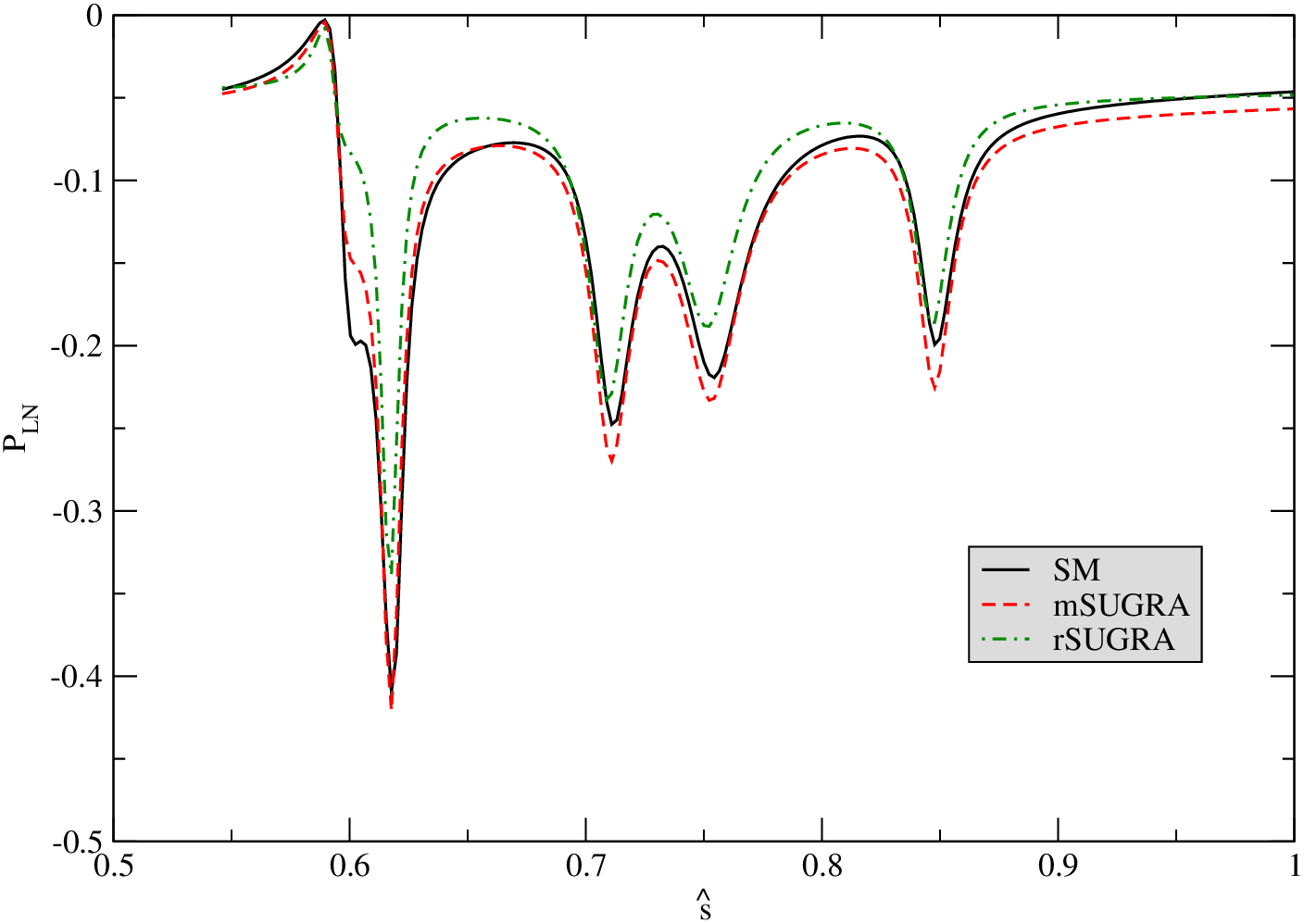}
\caption{\label{fig:pln_s}${\cal P}_{LN}$ with $\sh$ with all the
parameters same as in Fig.\ref{fig:dr_s}}
\end{figure}

In this section we will evaluate the double lepton polarization
asymmetries, i.e. where polarization of both the leptons is begin
measured. For this we have to define the polarization vectors of
$\ell^-$ and $\ell^+$. We will be going to use the convention as given
in many earlier works
\cite{Baek:2002wm,Xiong:2001up,RaiChoudhury:1999qb}. To evaluate the
polarized decay rates we  
have to introduce a spin projection operator defined by $N = 1/2 (1 +
\gamma_5 {\not S}_x)$ for $\ell^-$ and $M = 1/2 (1 + \gamma_5 {\not
W}_x)$ for $\ell^+$ where $x = L, N, T$ and corresponds to the
longitudinal, normal and transverse polarization asymmetries
respectively. Firstly we define the orthogonal unit vectors $S_x$ for
$\ell^-$ and $W_x$ for $\ell^+$ in rest frames of $\ell^-$ and
$\ell^+$ respectively as :
\beqa
S^\mu_L &\equiv& (0, \ev{L}) ~=~ \left(0, \frac{\pmv}{|\pmv|} \right)
                 \nonumber               \\
S^\mu_N &\equiv& (0, \ev{N}) ~=~ \left(0, \frac{\psv \times \pmv}{|\psv
             \times \pmv |}\right)
                 \nonumber               \\
S^\mu_T &\equiv& (0, \ev{T}) ~=~ \left(0, \ev{N} \times \ev{L}\right)
            \label{sec3:eq:1}             \\
W^\mu_L &\equiv& (0, \wv{L}) ~=~ \left(0, \frac{\ppv}{|\ppv|}\right)
                 \nonumber                \\
W^\mu_N &\equiv& (0, \wv{N}) ~=~ \left(0, \frac{\psv \times
                \ppv}{|\psv \times \ppv |} \right)
                 \nonumber                \\
W^\mu_T &\equiv& (0, \wv{T}) ~=~ (0, \wv{N} \times \wv{L})
                 \label{sec3:eq:2}
\eeqa
where $\pmv, \ppv$ and $\psv$ are the three momentas of $\ell^-,
\ell^+$ and strange (s) quark in center of mass frame (CM) frame of
$\ell^- \ell^+$ respectively. Now from the rest frame of respective
leptons we boost the four vectors $S_x$ and $W_x$ to the dilepton CM
frame. Only the longitudinal vectors, $S_L$ and $W_L$ will get boosted
by the Lorentz transformation to CM frame of $\ell^- \ell^+$ to a
value :
\beqa
S^\mu_L &=& \left( \frac{|\pmv|}{m_\ell}, \frac{E_\ell \pmv}{m_\ell
                  |\pmv|}   \right)
           \nonumber               \\
W^\mu_L &=& \left( \frac{|\pmv|}{m_\ell}, - \frac{E_\ell \pmv}{m_\ell
                |\pmv|} \right)
\label{sec3:eq:3}
\eeqa
where $E_\ell$ is the energy of any of the leptons (both have same
energy) in dileptonic CM frame. 

Now we can define the double polarization asymmetries as
\cite{Bensalem:2002ni} : 
\begin{widetext}
\beq
{\cal P}_{xy} \equiv  \frac{\left( \frac{d\Gamma( S_x, W_y )}{d\sh} -
                        \frac{d\Gamma( - S_x, W_y )}{d\sh} \right)
                       - \left( \frac{d\Gamma( S_x, - W_y )}{d\sh}
                         - \frac{d\Gamma(- S_x, - W_y )}{d\sh}\right)}
                      {\left( \frac{d\Gamma( S_x, W_y )}{d\sh} + 
                        \frac{d\Gamma( - S_x, W_y )}{d\sh} \right)
                       + \left( \frac{d\Gamma( S_x, - W_y )}{d\sh}
                         + \frac{d\Gamma( - S_x, - W_y )}{d\sh}\right)}
\label{sec3:eq:4}
\eeq
\end{widetext}
where the sub-index $x,y$ are $L,N$ or $T$.

We can get the expressions of double polarization asymmetries as :
\begin{widetext}
\beqa
{\cal P}_{LL} &=& \frac{1}{\bigtriangleup}
\Bigg[
- 4 \frac{(2 + \sh)}{\sh} \left(1 - \frac{2 \mlh^2}{\sh}\right) 
|\cseveff|^2
-  (1 + 2 \sh) \left(1 - \frac{2 \mlh^2}{\sh}\right) |\cneff|^2
- (1 - 8 \mlh^2 + 2 \sh - \frac{10 \mlh^2}{\sh}) |\cten|^2
\nonumber  \\
&& + {3 \over 2} (\sh -4 \mlh^2) |\cqone|^2
+ {3 \over 2} \sh |\cqtwo|^2    
- 12 \left(1 - \frac{2 \mlh^2}{\sh}\right) Re({\cseveff}^* \cneff)
+ 6 \mlh Re(\cten^* \cqtwo)
\Bigg]
\label{sec3:eq:5} \\
{\cal P}_{LN} &=& \frac{3 \pi}{2 \sqrt{\sh} \bigtriangleup}
\Bigg[
 2 \mlh  Im(\cten^* \cseveff)
- \sh Im({\cseveff}^* \cqtwo)
+  \mlh  Im(\cten^* \cneff)
- {1 \over 2} \sh Im({\cneff}^* \cqtwo)    \nonumber \\
&& + {1 \over 2} (\sh - 4 \mlh^2 ) Im(\cten^* \cqone)
\Bigg]
\label{sec3:eq:6}   \\
{\cal P}_{LT} &=& \frac{3 \pi}{2 \sqrt{\sh} \bigtriangleup}
\sqrt{1 - \frac{4 \mlh^2}{\sh}} 
\Bigg[ 
-  \mlh |\cten|^2
- 2 \mlh Re(\cten^* \cseveff )
+   \sh Re({\cseveff}^* \cqone) 
-  \mlh  \sh Re(\cten^* \cneff )  \nonumber \\
&& + {1 \over 2} \sh Re({\cneff}^* \cqone)
- {1 \over 2} \sh Re(\cten^* \cqtwo)
\Bigg]
\label{sec3:eq:7} \\
{\cal P}_{NL} &=& - {\cal P}_{LN}
\label{sec3:eq:8}  \\
{\cal P}_{NN} &=& \frac{1}{\bigtriangleup}
\Bigg[
- 4 \frac{(4 \mlh^2  - \sh + 2 \mlh^2  \sh + \sh^2 )}{\sh^2}
|\cseveff|^2   
+  (- 1 - 4 \mlh^2 + \sh - \frac{2 \mlh^2}{\sh} ) ( |\cneff|^2 
+ |\cten|^2 )  \nonumber
\\ 
&& - {3 \over 2} (\sh - 4 \mlh^2 ) |\cqone|^2
+ {3 \over 2} \sh |\cqtwo|^2      
 - 24 \frac{\mlh^2}{\sh}  Re({\cseveff}^* \cneff)
+ 6 \frac{\mlh}{\sh} Re(\cten^* \cqtwo)
\Bigg]
\label{sec3:eq:9}   \\ 
{\cal P}_{NT} &=&  \frac{2}{\bigtriangleup} \sqrt{1 - \frac{4 \mlh^2}{\sh}} 
\Bigg[
 (1 - \sh) Im(\cten^* \cneff )
+ 3 \mlh Im(\cten^* \cqone)
- {3 \over 2} \sh Im(\cqone^* \cqtwo)
\Bigg]
\label{sec3:eq:10}  \\
{\cal P}_{TL} &=& \frac{3 \pi}{2 \sqrt{\sh} \bigtriangleup}
\sqrt{1 - \frac{4 \mlh^2}{\sh}} 
\Bigg[
-  \mlh |\cten|^2
+ 2 \mlh Re(\cten^* \cseveff )
-  \sh Re({\cseveff}^* \cqone) 
+  \mlh \sh  Re(\cten^* \cneff )   \nonumber \\
&& - {1 \over 2} \sh  Re({\cneff}^* \cqone) 
- {1 \over 2} \sh Re(\cten^* \cqtwo)
\Bigg]
\label{sec3:eq:11}  \\
{\cal P}_{TN} &=& - {\cal P}_{NT}
\label{sec3:eq:12} \\
{\cal P}_{TT} &=& \frac{1}{\bigtriangleup} 
\Bigg[
- 4 \frac{(-4 \mlh^2  - \sh - 2 \mlh^2  \sh + \sh^2)}{\sh^2}
|\cseveff|^2 
+ (- 1 + 4 \mlh^2 + \sh + \frac{2 \mlh^2}{\sh} ) |\cneff|^2
+ (1 + 4 \mlh^2 - \sh - \frac{10 \mlh^2}{\sh}) |\cten|^2
       \nonumber \\
&& + {3 \over 2} (\sh - 4 \mlh^2) |\cqone|^2
- {3 \over 2} \sh |\cqtwo|^2   
 + 24 \frac{\mlh^2}{\sh} Re({\cseveff}^* \cneff )
- 6 \frac{\mlh}{\sh} Re(\cten^* \cqtwo)
\Bigg]
\label{sec3:eq:13}
\eeqa
\end{widetext}
where $\bigtriangleup$ is given in eqn.(\ref{sec2:eq:9}).


\section{\label{section:4} Numerical analysis, Results and discussion}

In this section we will be going to discuss our numerical analysis and
the results of our numerical analysis. Firstly we will be going to
list the SM values of the branching ratio of \btoxstt and various
double polarization asymmetries we have presented in previous sections
in Table \ref{sec4:tab:1}.

\begin{table*}
\begin{ruledtabular}
\begin{tabular}{c | c  c  c  c  c  c  c  c  c }  
\hspace{0.2cm} Br(\btoxstt)  \hspace{.2cm}
& \hspace{.2cm} ${\cal P}_{LL}$ \hspace{.2cm}
& \hspace{.2cm} ${\cal P}_{LN}$ \hspace{.2cm}  
& \hspace{.2cm} ${\cal P}_{LT}$ \hspace{.2cm}
& \hspace{.2cm} ${\cal P}_{NL}$ \hspace{.2cm}  
& \hspace{.2cm} ${\cal P}_{NN}$ \hspace{.2cm}
& \hspace{.2cm} ${\cal P}_{NT}$ \hspace{.2cm} 
& \hspace{.2cm} ${\cal P}_{TL}$ \hspace{.2cm}
& \hspace{.2cm} ${\cal P}_{TN}$ \hspace{.2cm}
& \hspace{.2cm} ${\cal P}_{TT}$ \hspace{.2cm} \\  \hline 
$1.29 \times 10^{-7}$ 
& - 0.082  & - 0.137   & - 0.136
&   0.136  &   0.122   & - 0.017
& - 0.397  &   0.017   &  -0.216     \\  
\end{tabular}
\caption{Standard Model predictions of the observables for \btoxstt}
\label{sec4:tab:1}
\end{ruledtabular}
\end{table*}

We have analyzed the SUSY effects on the various observables which we
have listed in previous section. MSSM is although the simplest and the
one having least number of parameters, extension of the SM. But still
it has large number of parameters which makes it difficult to do
phenomenology with it. But we do have some models, like Dilaton,
moduli, mSUGRA, \dots etc, which reduces this vast parameter space to
a manageable level. In our numerical work we will be going to use one
of the more popular unified model called Supergravity (SUGRA)
model. The main feature of SUGRA (we will call this as minimal SUGRA
or mSUGRA) model is that a unification of all the scalars, fermions
and coupling constants is assumed at GUT scale. So effectively in
mSUGRA framework we have five parameters which are : $m$ (unified mass
of all the scalars), $M$ (unified mass of all the gauginos), $A$
(universal trilinear coupling constant), $tan\beta$ (ratio of vev of
the two Higgs doublets) and $sgn(\mu)$\footnote{the convention about
$sgn(\mu)$ we will be going to follow is that it appears in chargino
mass matrix with a +ve sign}. 

\begin{figure}
\includegraphics[width=3in]{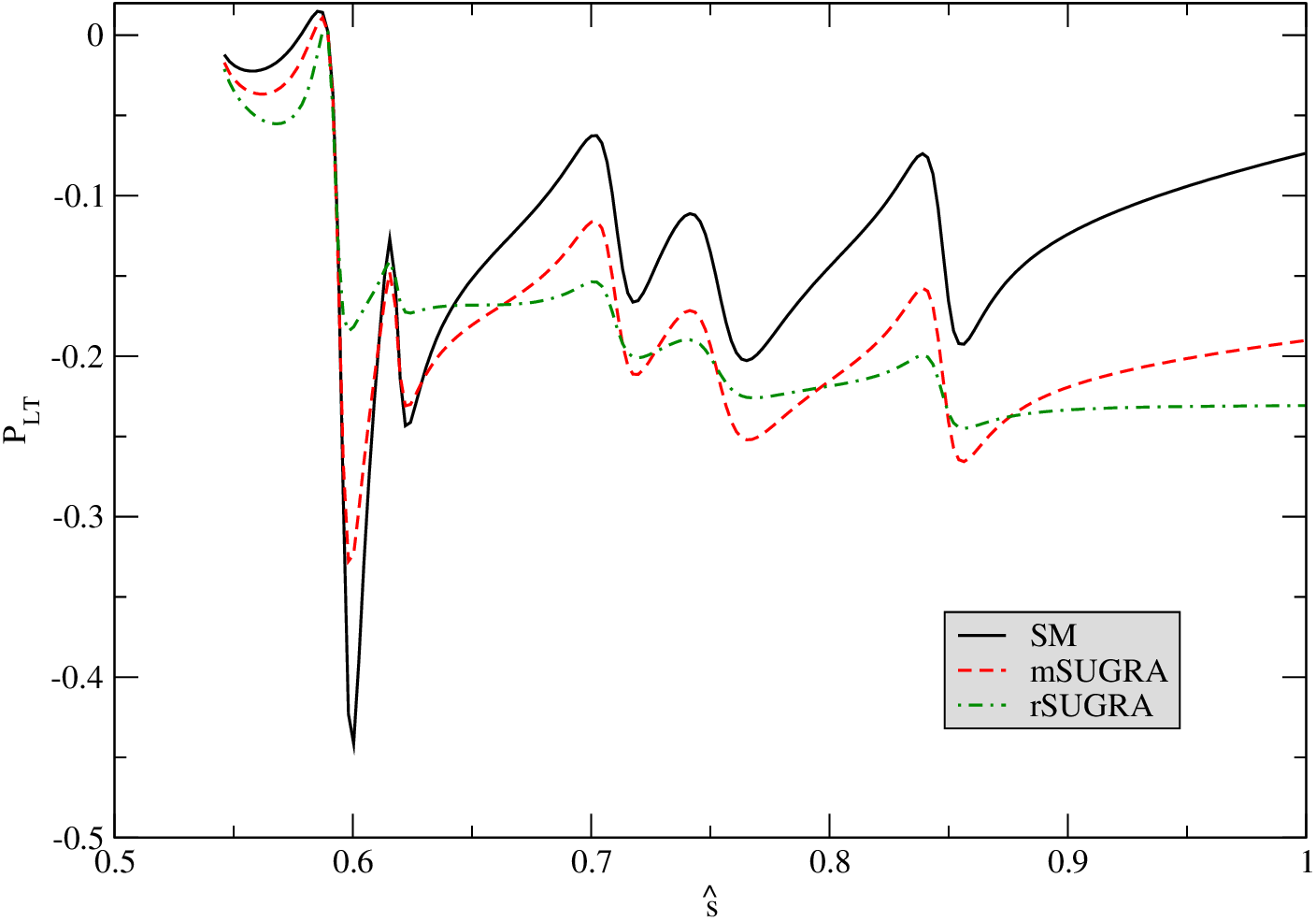}
\caption{\label{fig:plt_s}${\cal P}_{LT}$ with $\sh$ with all the
parameters same as in Fig.\ref{fig:dr_s}}
\end{figure}
\begin{figure}
\includegraphics[width=3in]{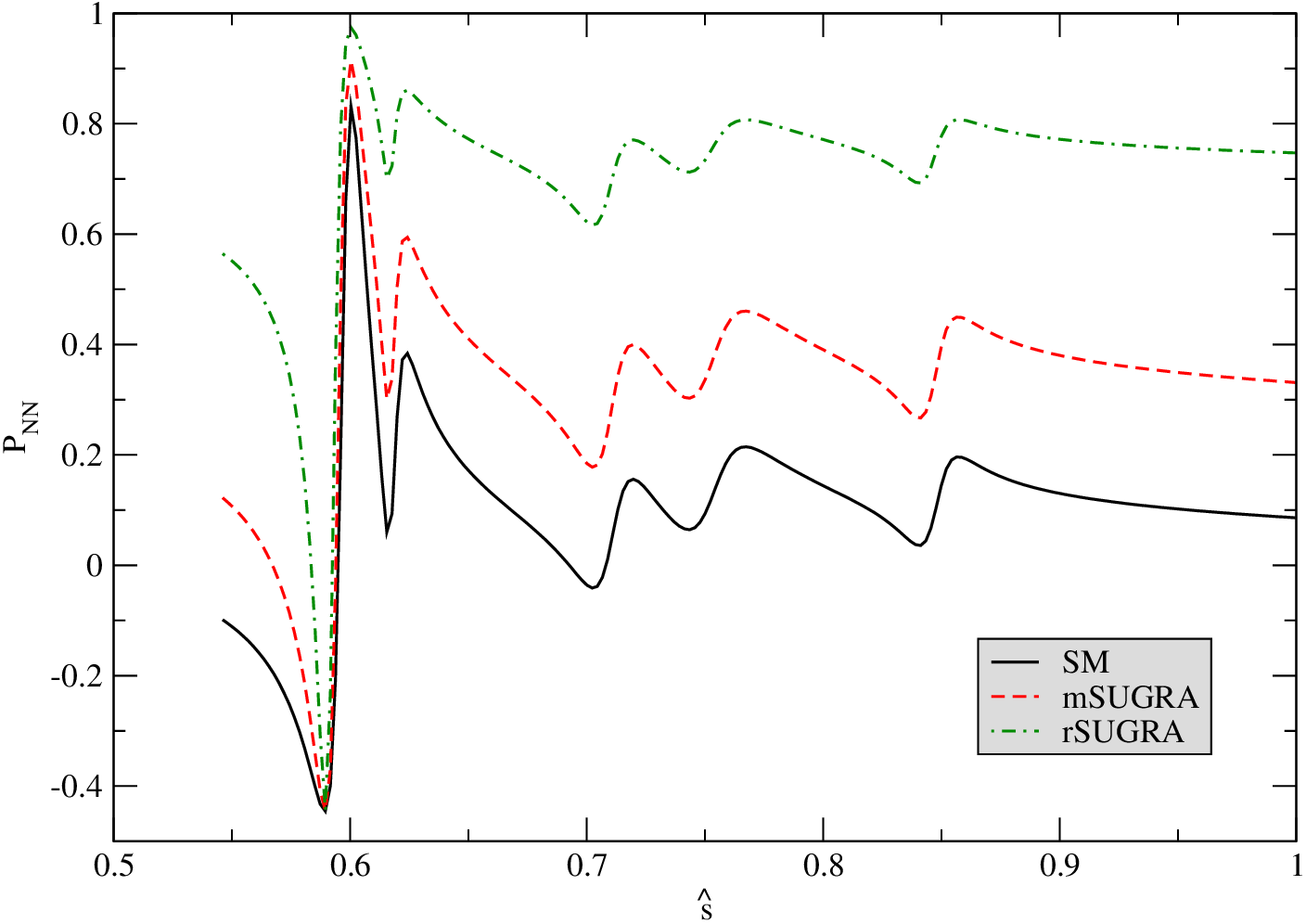}
\caption{\label{fig:pnn_s}${\cal P}_{NN}$ with $\sh$ with all the
parameters same as in Fig.\ref{fig:dr_s}} 
\end{figure}
\begin{figure}
\vskip .6cm
\includegraphics[width=3in]{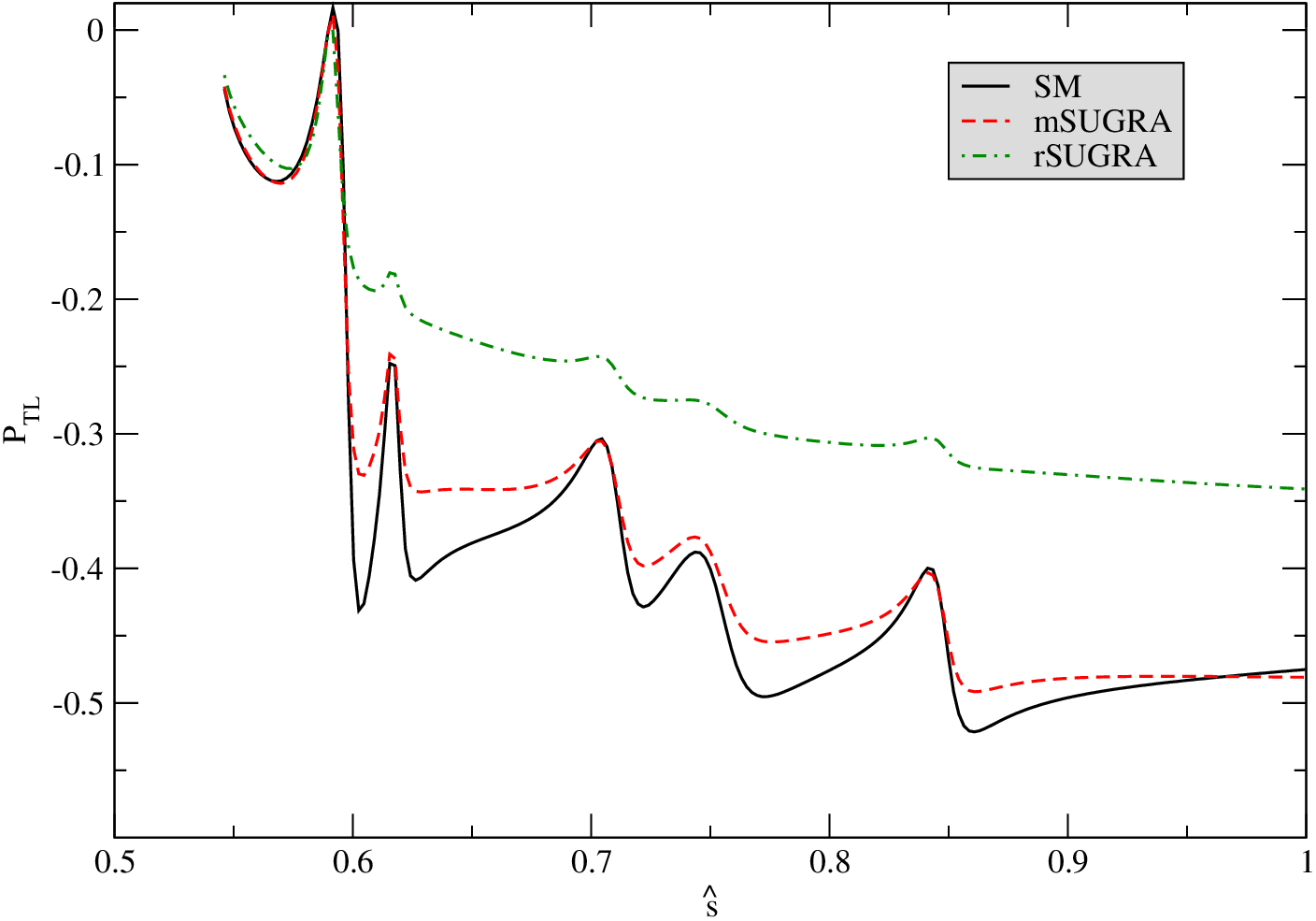}
\caption{\label{fig:ptl_s}${\cal P}_{TL}$ with $\sh$ with all the 
parameters same as in Fig.\ref{fig:dr_s}} 
\end{figure}
\begin{figure}
\vskip .6cm
\includegraphics[width=3in]{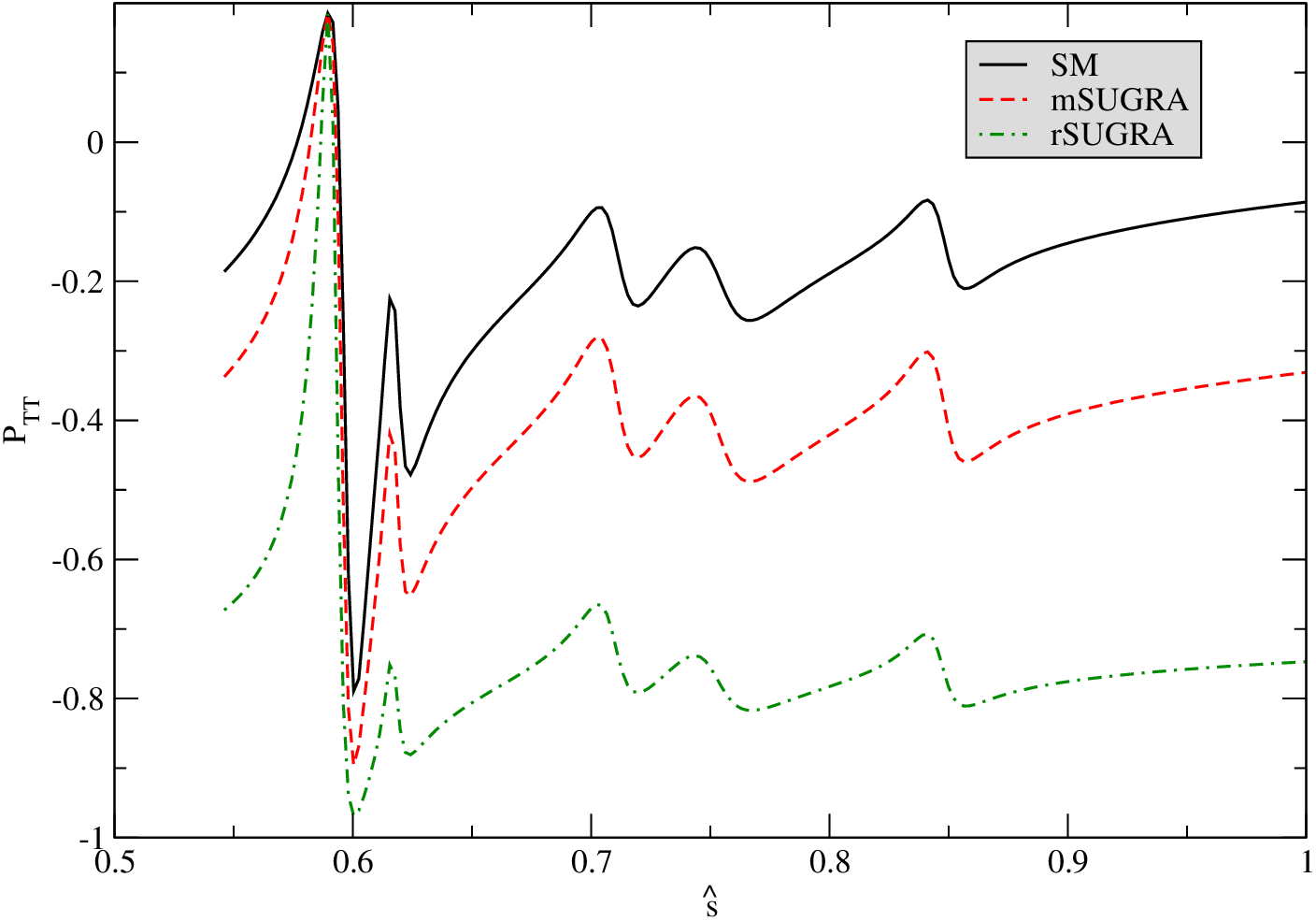}
\caption{\label{fig:ptt_s}${\cal P}_{TT}$ with $\sh$ with all the
parameters same as in Fig.\ref{fig:dr_s}} 
\end{figure}

\par But as we know that universality of scalar masses at GUT scale is
not a necessary condition of SUGRA \cite{goto1}. One can in principle
have a different unified mass of the squark sector and Higgs sector at
GUT scale. We will explore this scenario also, which we will call as 
relaxed SUGRA (rSUGRA) model. The additional parameter which we will
be going to have in this model we will take as mass of pseudo-scalar
Higgs boson ($m_A$).  

\par We will be going to work in the high $tan\beta$ region of the
SUGRA parameter space because its only in this region the effects of
NHBs becomes more prominent. There are some constraints on the MSSM
parameter space from experimental observations of $B \to X_s \gamma$
\cite{expbsg}. In our analysis we will be going to consider only that region
of parameter space which satisfy the 95\% CL bound \cite{expbsg} :
$$ 
2 \times 10^{-4} < Br(B \to X_s \gamma) < 4.5 \times 10^{-4}
$$
which is agreement with CLEO and ALEPH results.

We will also present the numerical results of the average polarization
asymmetries. The averaging procedure which we will be going to use is
defined as :
\beq
\langle {\cal P} \rangle \equiv \frac{\displaystyle{\int_{(3.646 +
0.02)^2/m_B^2}^{(m_B - m_{K^*})^2/m_B^2}} {\cal P} \frac{d \Gamma}{d
\sh} d \sh}{ \displaystyle{\int_{(3.646 +
0.02)^2/m_B^2}^{(m_B - m_{K^*})^2/m_B^2}} \frac{d \Gamma}{d
\sh} d \sh} 
\label{sec4:eq:1}
\eeq
which means in integrating the observables we will be going to
consider the region of dilepton invariant mass which is above the
first charmonium resonance (after the threshold of the process $b \to
s \tau^+ \tau^-$). We will be not be going to present the result of
${\cal P}_{NT}$ (and ${\cal P}_{TN}$) as they are very small.

\par We have presented the plots, of the various kinematical
variables, we presented in sections \ref{section:2} and
\ref{section:3}, with the (scaled) invariant mass of the dilepton. In
Fig.\ref{fig:dr_s} we have shown the variation of the branching ratio
of \btoxstt with dilepton invariant mass in three different models we
have considered, namely SM, mSUGRA and rSUGRA. In Fig. \ref{fig:pll_s},
\ref{fig:pln_s}, \ref{fig:plt_s}, \ref{fig:pnn_s}, \ref{fig:ptl_s} and
\ref{fig:ptt_s} we have plotted the distributions of various double
polarization asymmetries with dileptonic invariant mass. As we can see
from these figures that SUSY can change these distributions
substantially over the whole kinematically allowed region. We have
done the detailed scanning of the mSUGRA and rSUGRA parameter space
and have presented the results of the integrated polarization
asymmetries, as defined in eqn.(\ref{sec4:eq:1}). In
Fig. \ref{fig:pll_tb}, \ref{fig:pln_tb}, \ref{fig:plt_tb},
\ref{fig:pnn_tb}, \ref{fig:ptl_tb} and \ref{fig:ptt_tb} we have shown
the plots of various integrated polarization asymmetries with
$tan\beta$ for various values of $M$ (the unified gaugino mass at GUT
scale) in the mSUGRA model. As we can see that there is a substantial
changes in the SM and mSUGRA model prediction. Specially for ${\cal
P}_{LL}$ where depending upon the mSUGRA parameters it can even change
its sign. In Fig. \ref{fig:pll_ma}, \ref{fig:pln_ma},
\ref{fig:plt_ma}, \ref{fig:pnn_ma}, \ref{fig:ptl_ma} and
\ref{fig:ptt_ma} we have plotted various integrated polarization
asymmetries as function $m_A$ (mass of pseudo-scalar Higgs boson) for
various values of $tan\beta$. There also one can observe major changes
in the predictions of various integrated polarization asymmetries as
compared to SM values. 

As has already been mentioned in many works
\cite{Baek:2002wm,RaiChoudhury:1999qb,Kruger:1996cv,Bensalem:2002ni,Choudhury:2003a}
that polarization  
asymmetries are useful in finding out the structure of the effective
Hamiltonian and hence the physics underlying it. As also been argued
by Bensalam et al. \cite{Bensalem:2002ni}, if we work within the SM
then for inclusive decay 
\btoxsll we have five theoretical parameters which are : four Wilsons
($C_7$, $C_{10}$, real and imaginary part of $\cneff$) and $m_b$,
they can in principle be completely determined using the three
$\tau^-$ polarization asymmetries, the total decay rate and the FB
asymmetry i.e. five observables. But as known from earlier results
\cite{RaiChoudhury:1999qb,Kruger:1996cv} the normal polarization
asymmetry is very small and hence we 
have to consider the polarization asymmetry of $\tau^+$ also. Also the
measurement of FB asymmetry requires b-tagging. So if we have a
untagged sample then the observables which we have are : decay rate,
two polarization asymmetries of $\tau^-$ (we are neglecting the normal
one being very small) and two polarization asymmetries of $\tau^+$
(again neglecting the normal one). So we have five parameters and five
observables which should in principle be sufficient. But there is no
other constraint which will give us the cross-check of SM. So along
these arguments Bensalam et.al. constructed double polarization
asymmetries which give us large number of observables to have a good
cross-checking of SM.

Let's now examine the situation if we believe in SUSY extension
the SM. As has been argued in many works
\cite{Choudhury:1999ze,Baek:2002wm,Xiong:2001up} that if we consider 
the decay channel $b \to s \tau^+ \tau^-$ in SUSY extension of SM
\footnote{in fact this is true even if we consider the two Higgs
Doublet model (2HDM) extension of SM \cite{Skiba:1993mg} },
then we have to extend the SM list of operators by introduction of two
new operators $\cqone$ and $\cqtwo$. So now our theoretical parameters
would be seven (five SM one and two new Wilson coefficients). So we
require more number of observables to fix up this new structure of
effective Hamiltonian and hence the double polarization asymmetries
could possibly be very useful. 

\par The model which we are considering is MSSM. In MSSM if we assume
the MSSM parameters to be real then the only source of CP violation is
the CKM matrix. 
The transition \btosll is a CP conserving process. Let's try to
analyze the number of additional observables which we now have
\footnote{by additional observables we mean the observables due to
measurement of double polarization asymmetries} . For this we consider
two possibilities : one where we can have {\sl b-tagging} and the
other where there is no {\sl b-tagging}. If we say that the double
polarization asymmetries for \btoxstt are ${\cal P}_{ij}$ then for the
conjugated process ${\bar B} \to {\bar X}_s \tau^+ \tau^-$ we denote
them with ${\bar {\cal P}}_{ij}$. But in general ${\bar {\cal P}}_{ij}
= \pm {\cal P}_{ij}$\footnote{in fact except for ${\cal P}_{NL}$ and
${\cal P}_{NT}$ the sign is always positive. For  ${\cal P}_{NL}$ and
${\cal P}_{NT}$ the sign is negative}. 

\par So if we consider the first case where there is no {\sl
b-tagging}. In a untagged sample we can measure four additional
observables namely ${\cal P}_{LL}, {\cal P}_{NN}, {\cal P}_{TT}$ and
(${\cal P}_{LT} + {\cal P}_{TL}$). 

\par But if we consider that we can tag $b$ in that case we have all
the nine double polarization asymmetries available to us which will
give us very useful probes into the structure of effective
Hamiltonian. 


\begin{acknowledgments}
I am thankful to Prof. S. Rai Choudhury for useful discussions and
comments during the course of the work. 
Thanks are due to
Prof. D. London for initiation of this work which is the extension of
his earlier work \cite{Bensalem:2002ni}. 
I am also thankful to
Dr. Ashok Goyal for carefully reading the manuscript. This work is
supported under SERC scheme of Department of Science and Technology
(DST), India.  
\end{acknowledgments}


\appendix

\section{\label{appendix:1} Input parameters}

\begin{center}
$m_B ~=~ 5.26$ GeV \ , \ $m_b ~=~ 4.8$ GeV  \ , \
$ m_c ~=~ 1.4 $ GeV  \\
$m_\tau ~=~ 1.77$  GeV  \ , \
$m_w ~=~ 80.4$  GeV  \ , \\ 
$m_z ~=~ 91.19$  GeV  \ , \
$V_{tb} V^*_{ts} = 0.0385$  , \\ 
$\alpha = {1 \over 129}$  \ , \
 $\Gamma_B = 4.22 \times 10^{-13}$ GeV  \\
$G_F = 1.17 \times 10^{-5} ~{\rm GeV}^{-2}$ 
\end{center}


\newpage


\begin{widetext}
\begin{figure*}
\centering{
\includegraphics[width=3in]{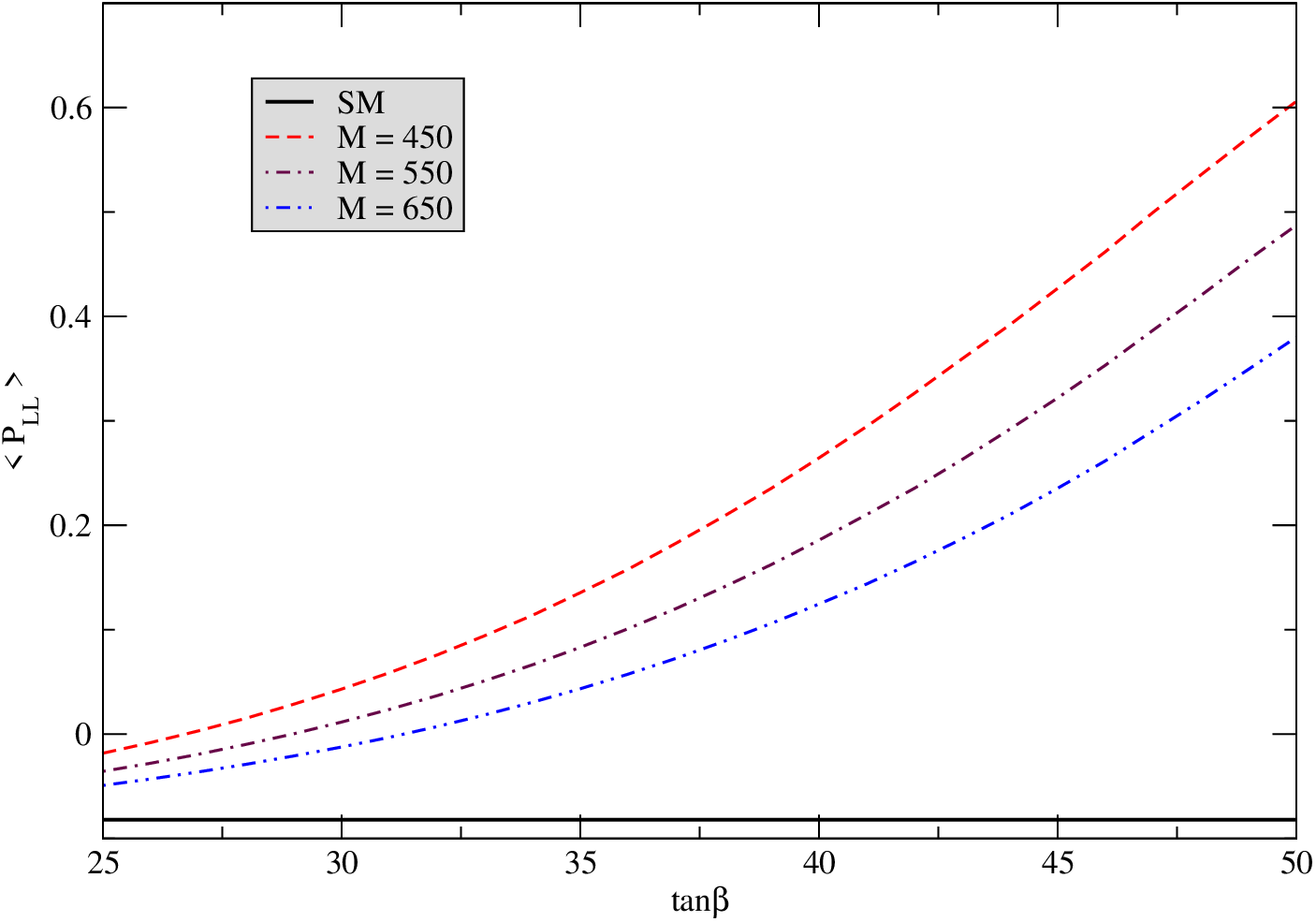}
\caption{\label{fig:pll_tb} Integrated ${\cal P}_{LL}$ with $tan\beta$ 
in mSUGRA model. Parameters same as given in Fig.\ref{fig:dr_tb}}}
\vskip 1cm
\includegraphics[width=3in]{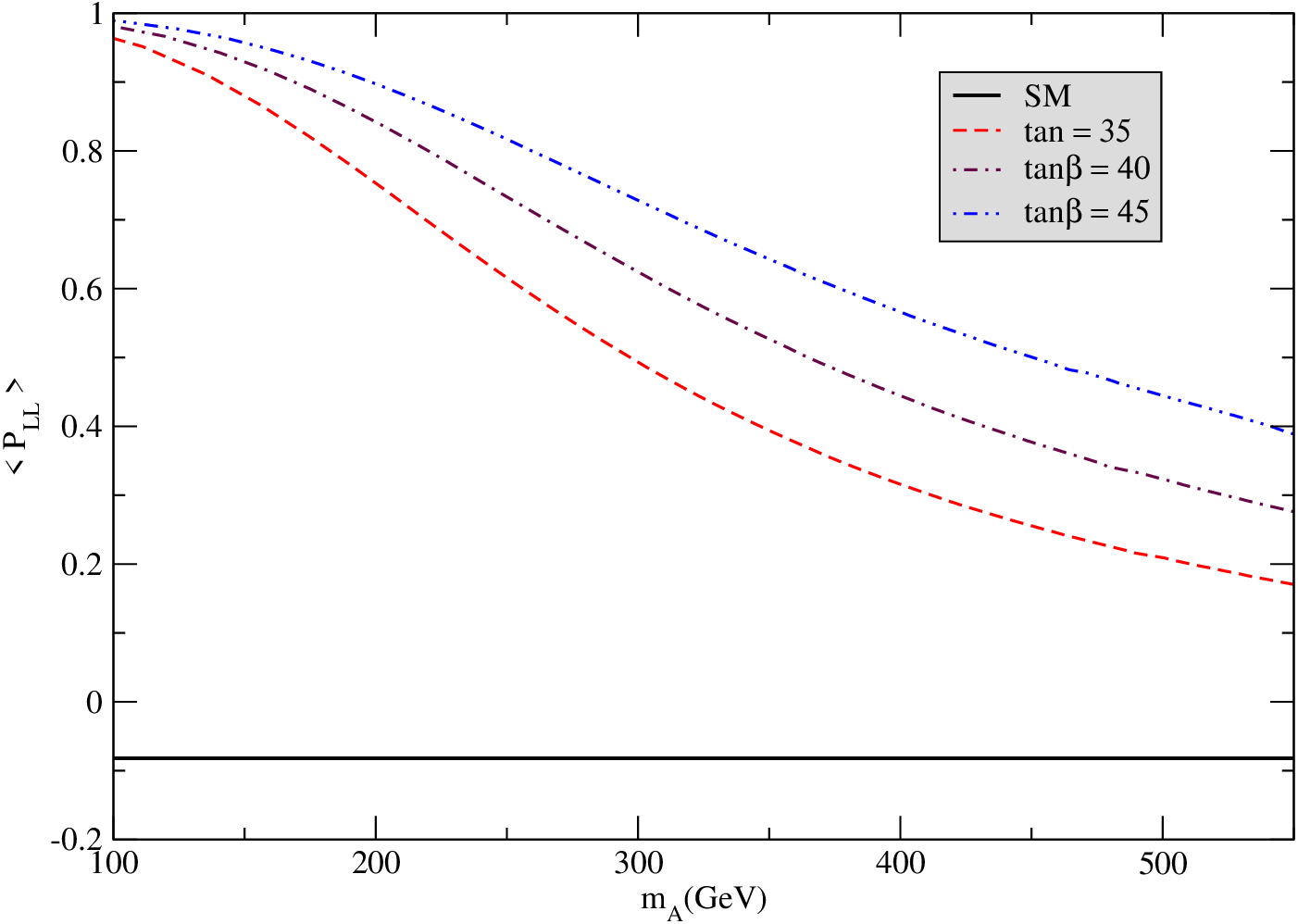}
\caption{\label{fig:pll_ma}Integrated ${\cal P}_{LL}$ with $m_A$GeV in
rSUGRA model. Parameters same as given in Fig. \ref{fig:dr_ma}}
\end{figure*}
\begin{figure*}
\vskip .5cm
\includegraphics[width=3in]{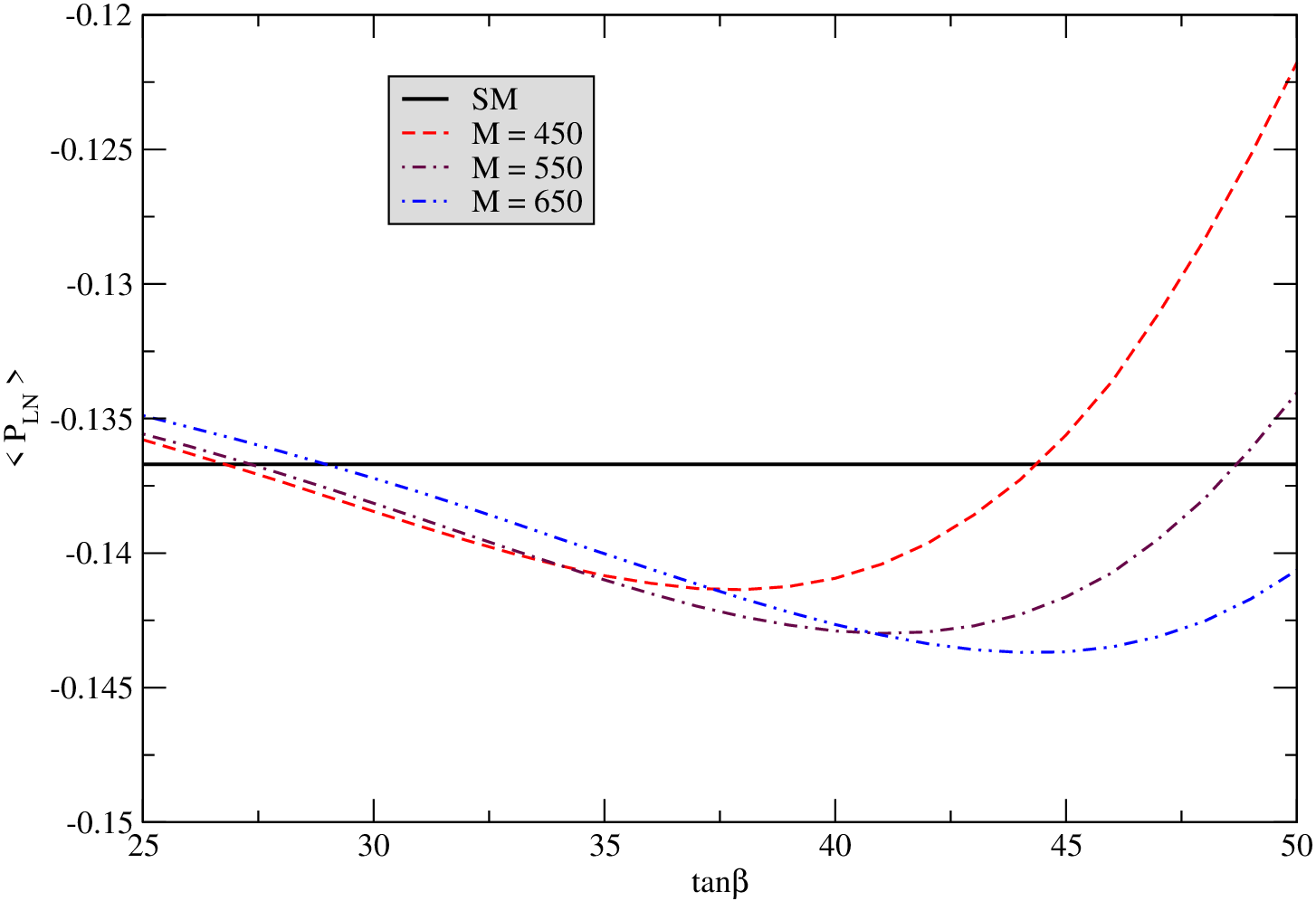}
\caption{\label{fig:pln_tb}Integrated ${\cal P}_{LN}$ with $tan\beta$
in mSUGRA model. Parameters same as given in Fig.\ref{fig:dr_tb}}
\end{figure*}
\begin{figure*}
\includegraphics[width=3in]{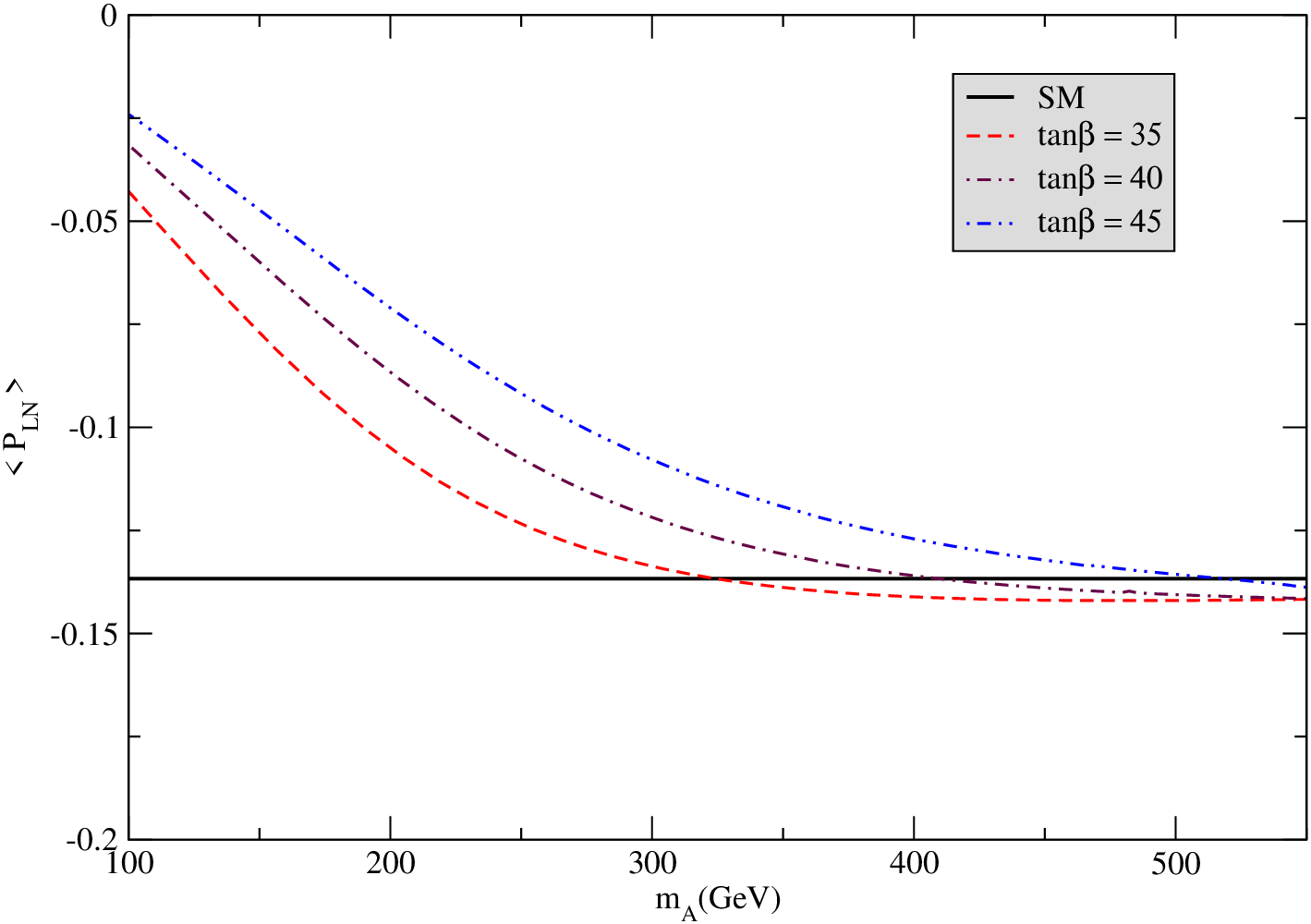}
\caption{\label{fig:pln_ma}Integrated ${\cal P}_{LN}$ with $m_A$GeV in
rSUGRA model. Parameters same as given in Fig.\ref{fig:dr_ma}}
\end{figure*}
\begin{figure*}
\includegraphics[width=3in]{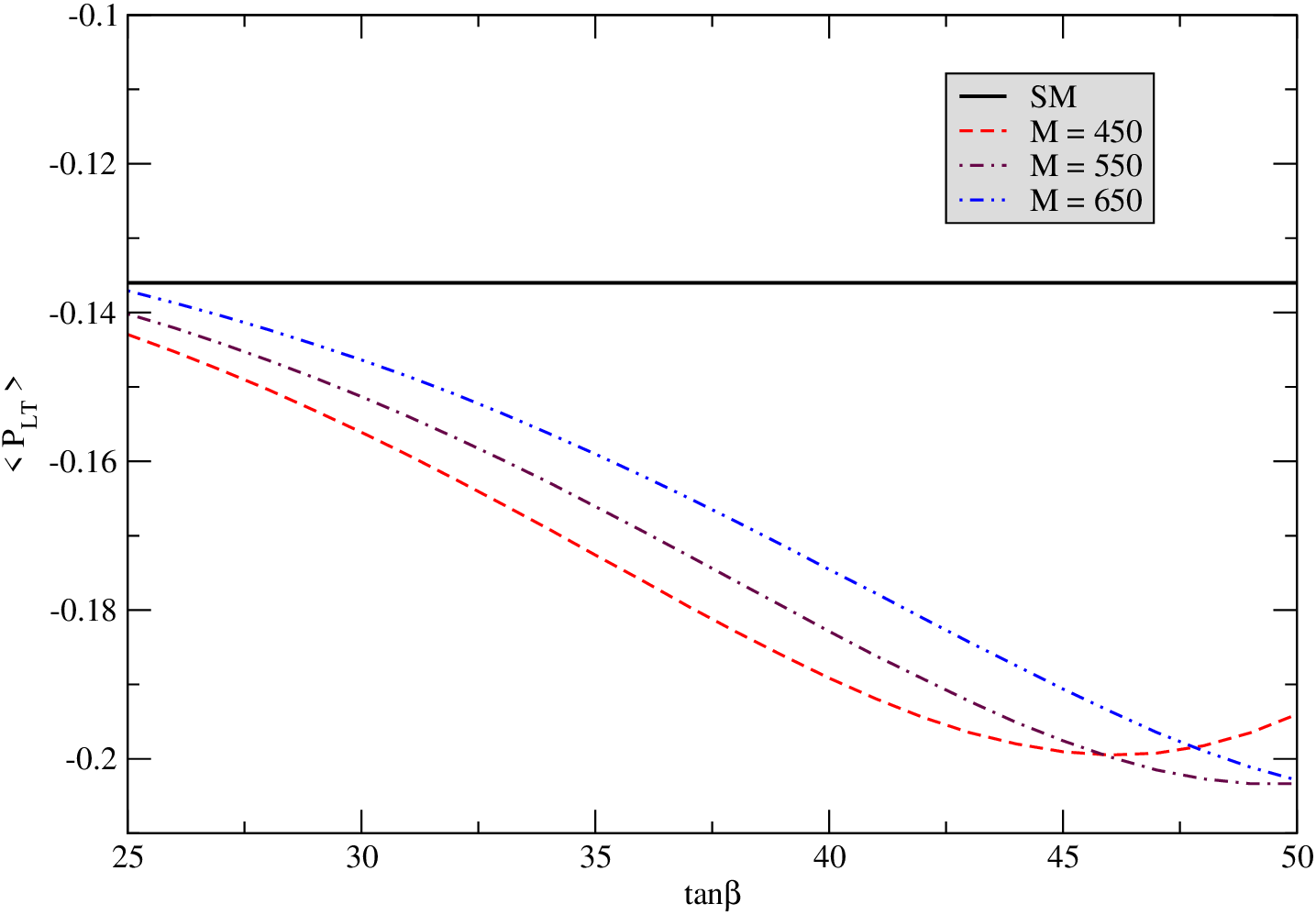}
\caption{\label{fig:plt_tb}Integrated ${\cal P}_{LT}$ with $tan\beta$
in mSUGRA model. Parameters same as given in Fig. \ref{fig:dr_tb}}
\end{figure*}
\begin{figure*}
\includegraphics[width=3in]{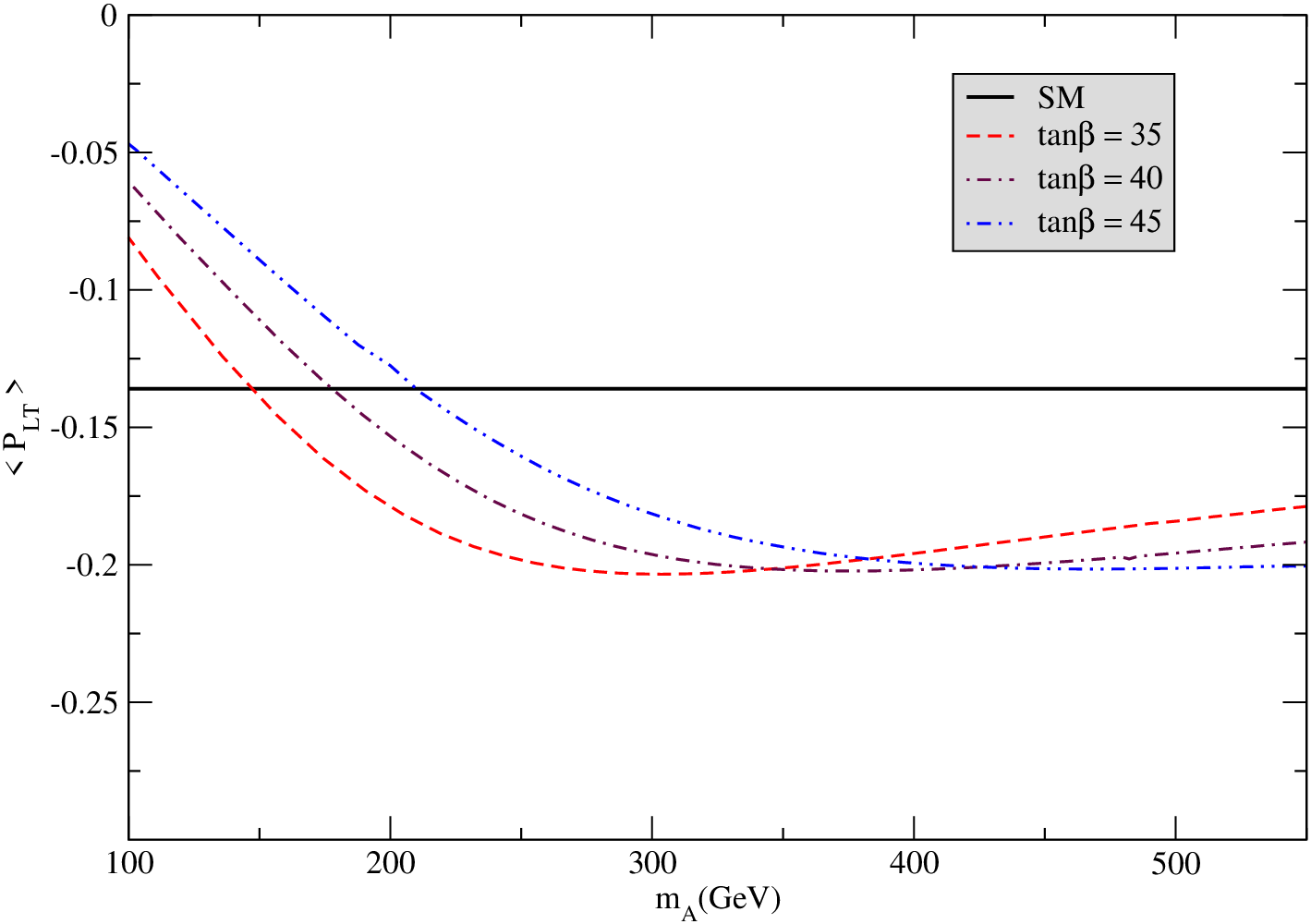}
\caption{\label{fig:plt_ma}Integrated ${\cal P}_{LT}$ with $m_A$GeV in
rSUGRA model. Parameters same as given in Fig. \ref{fig:dr_ma}}
\end{figure*}
\begin{figure*}
\includegraphics[width=3in]{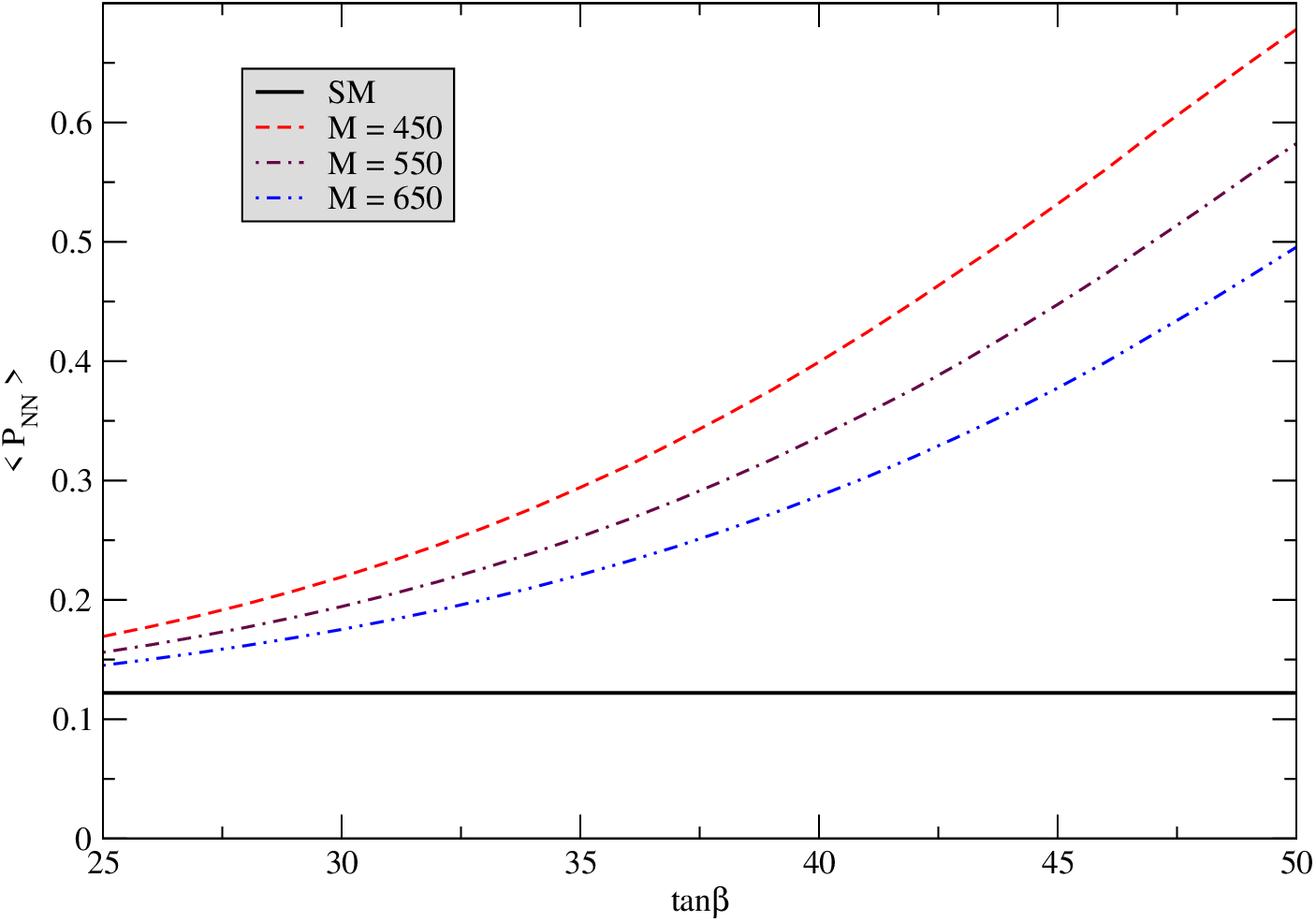}
\caption{\label{fig:pnn_tb}Integrated ${\cal P}_{NN}$ with $tan\beta$
in mSUGRA model. Parameters same as given in Fig. \ref{fig:dr_tb}}
\end{figure*}
\begin{figure*}
\includegraphics[width=3in]{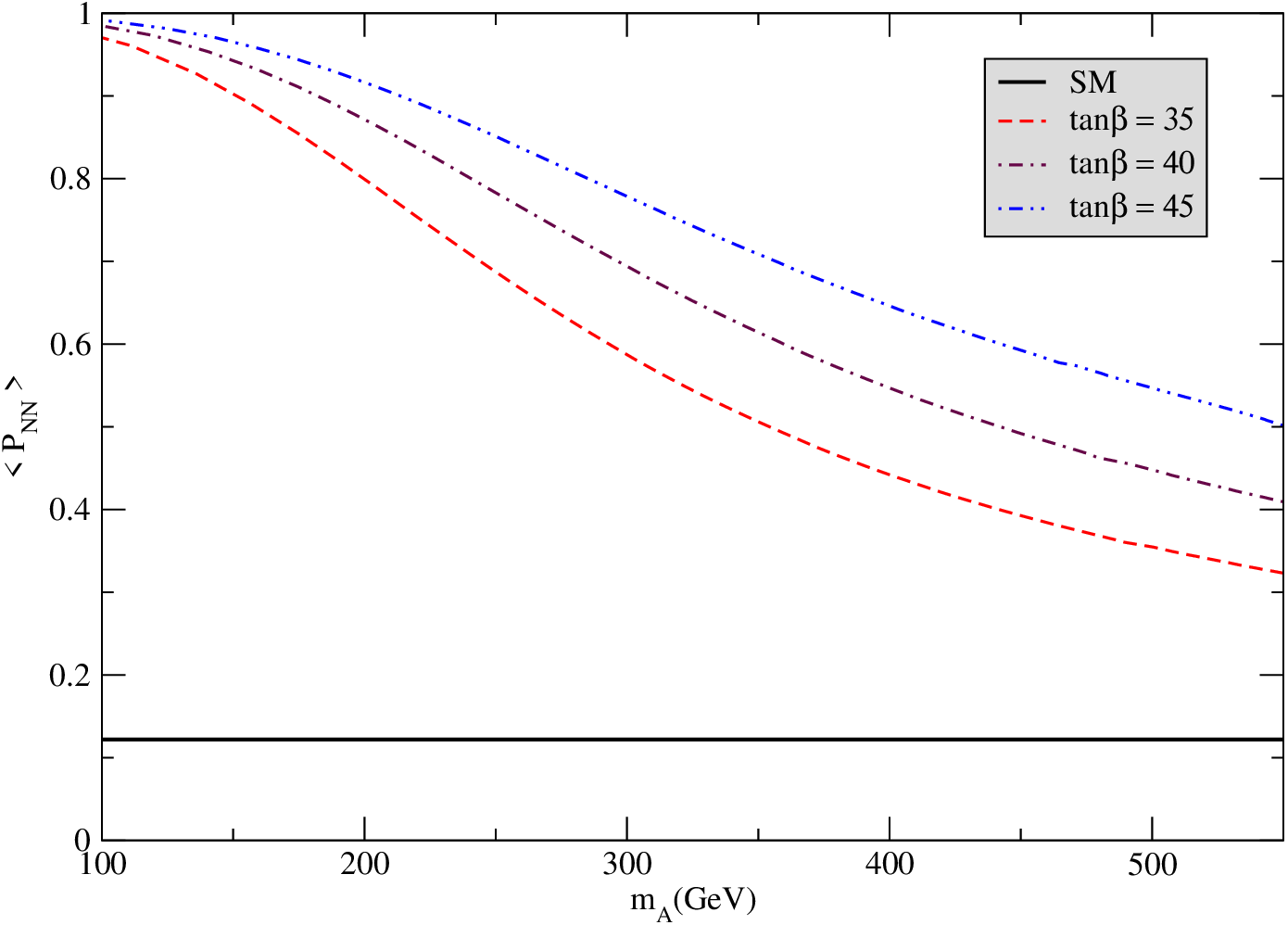}
\caption{\label{fig:pnn_ma}Integrated ${\cal P}_{NN}$ with $m_A$GeV in
rSUGRA model. Parameters same as given in Fig. \ref{fig:dr_ma}}
\end{figure*}
\begin{figure*}
\includegraphics[width=3 in]{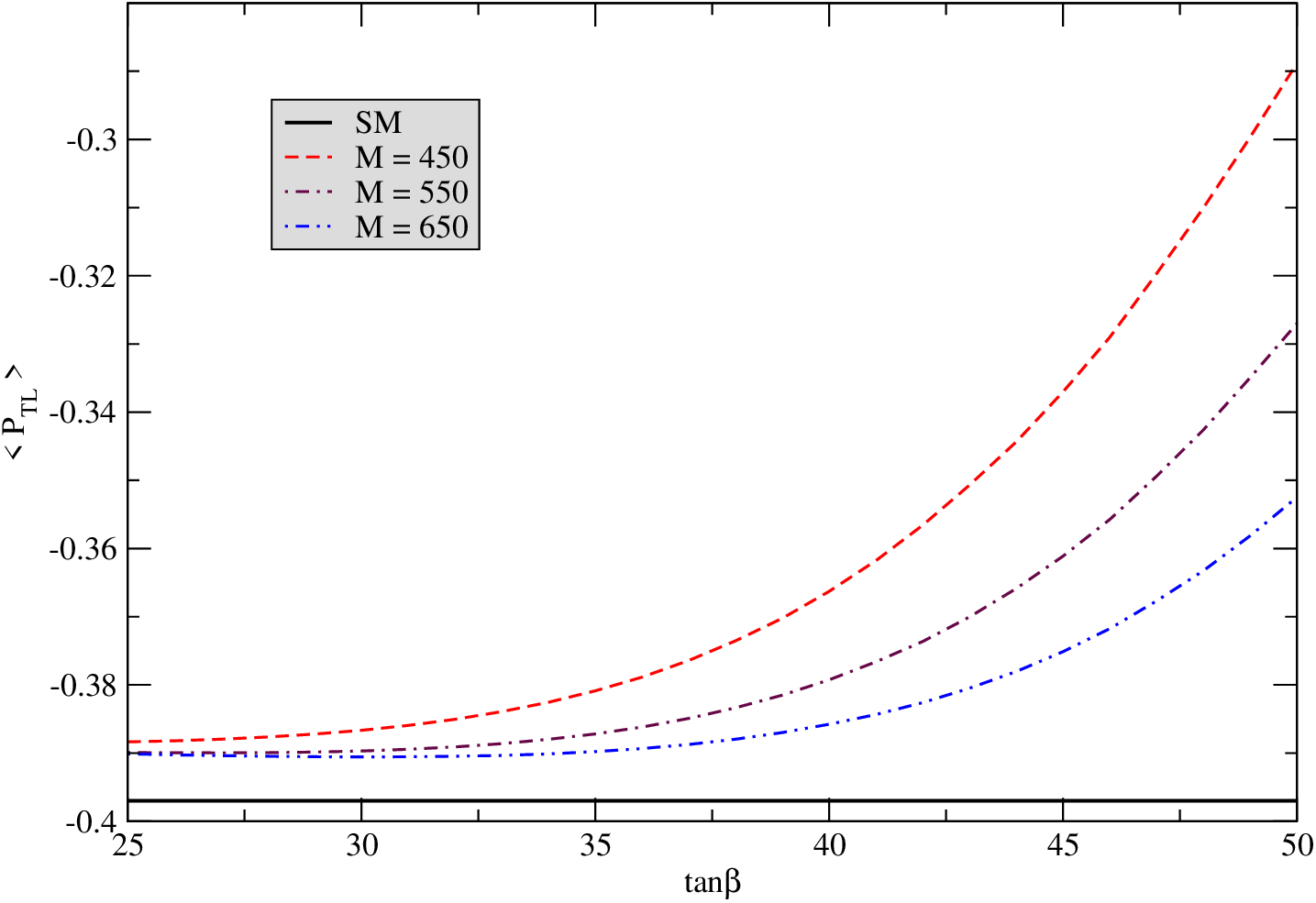}
\caption{\label{fig:ptl_tb}Integrated ${\cal P}_{TL}$ with $tan\beta$
in mSUGRA model. Parameters same as given in Fig. \ref{fig:dr_tb}}
\end{figure*}
\begin{figure*}
\includegraphics[width=3 in]{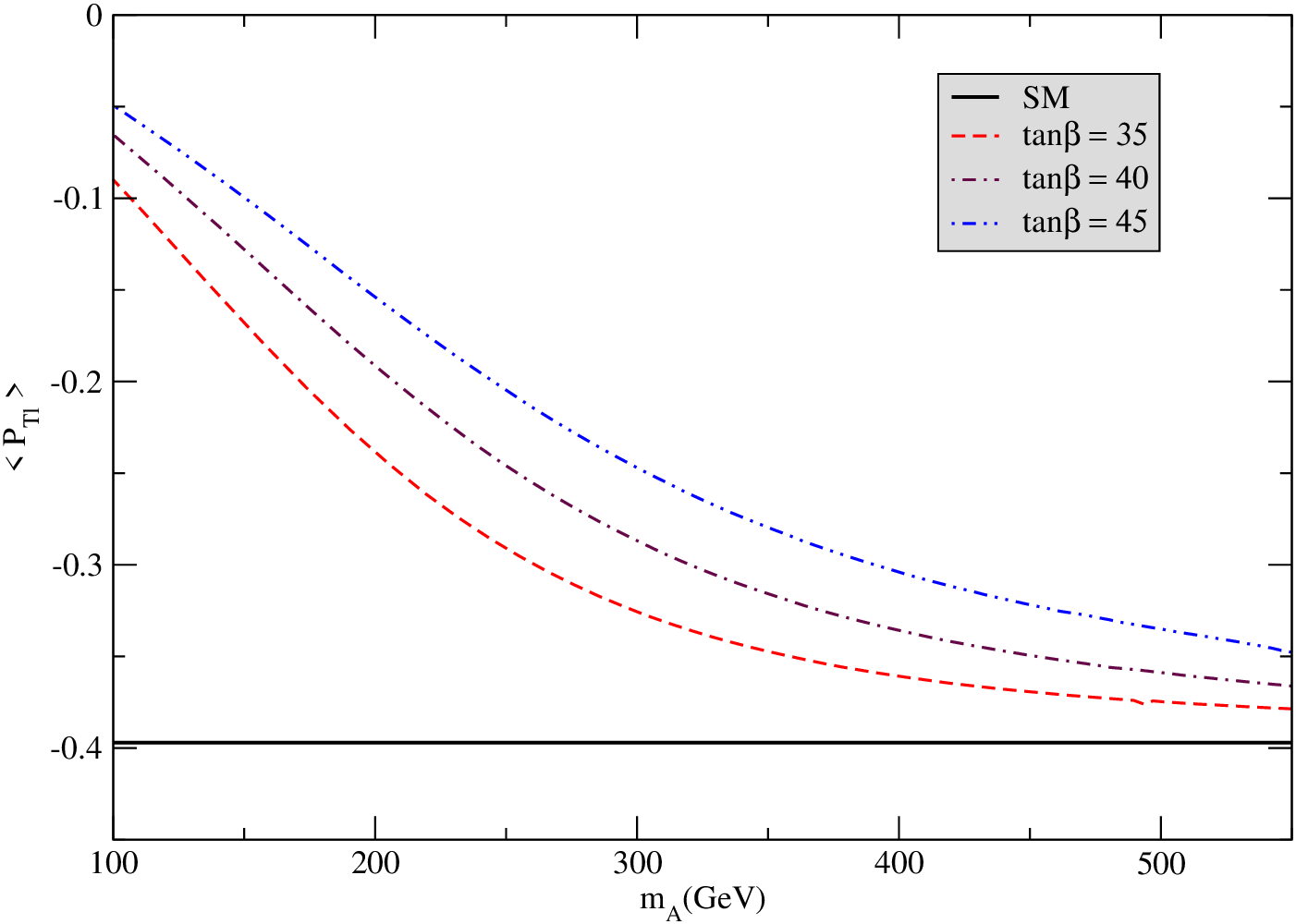}
\caption{\label{fig:ptl_ma}Integrated ${\cal P}_{TL}$ with $m_A$GeV in
rSUGRA model. Parameters same as given in Fig. \ref{fig:dr_ma}}
\end{figure*}
\begin{figure*}
\includegraphics[width=3 in]{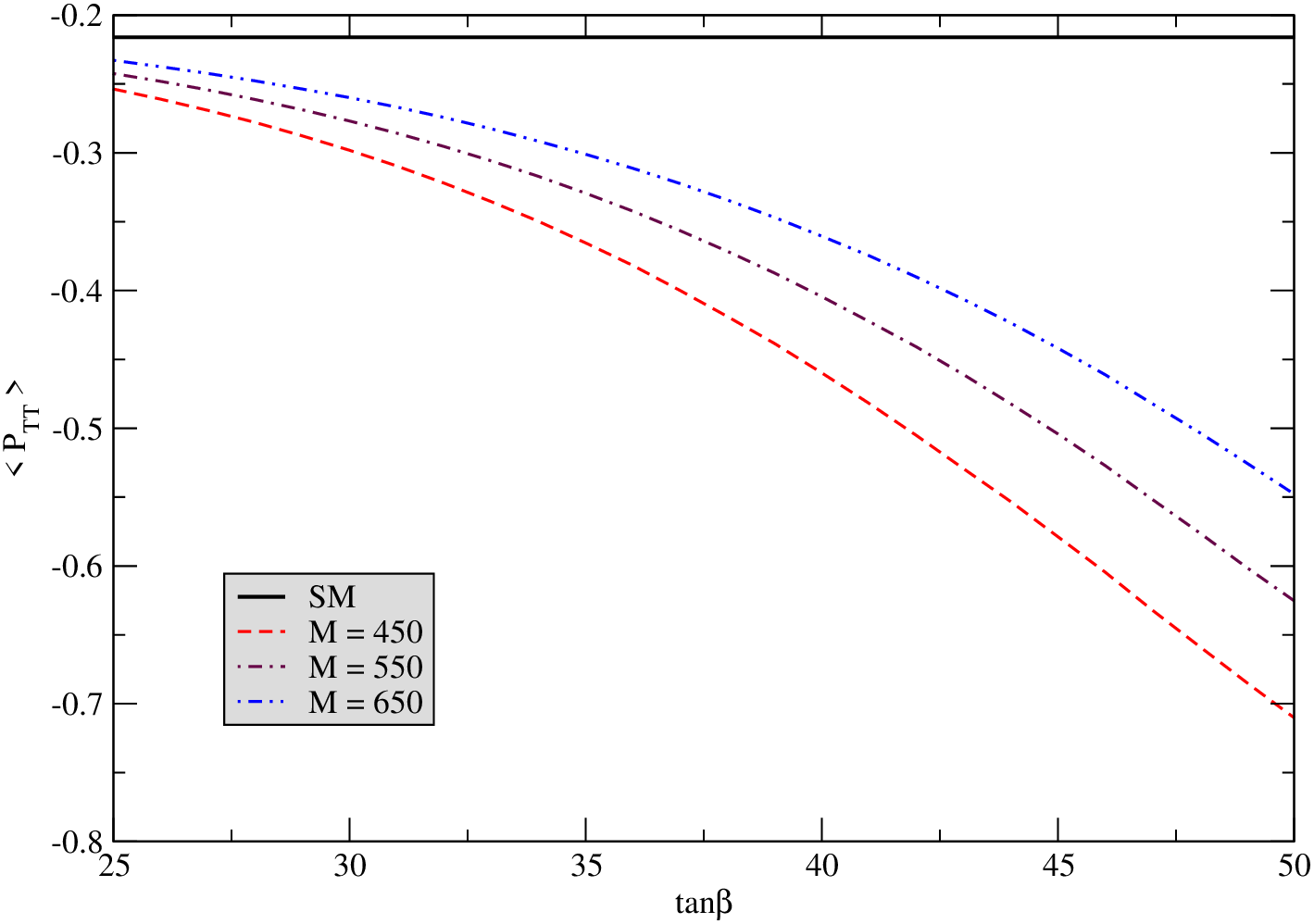}
\caption{\label{fig:ptt_tb}Integrated ${\cal P}_{TT}$ with $tan\beta$
in mSUGRA model. Parameters same as given in Fig. \ref{fig:dr_tb}}
\end{figure*}
\begin{figure*}
\includegraphics[width=3 in]{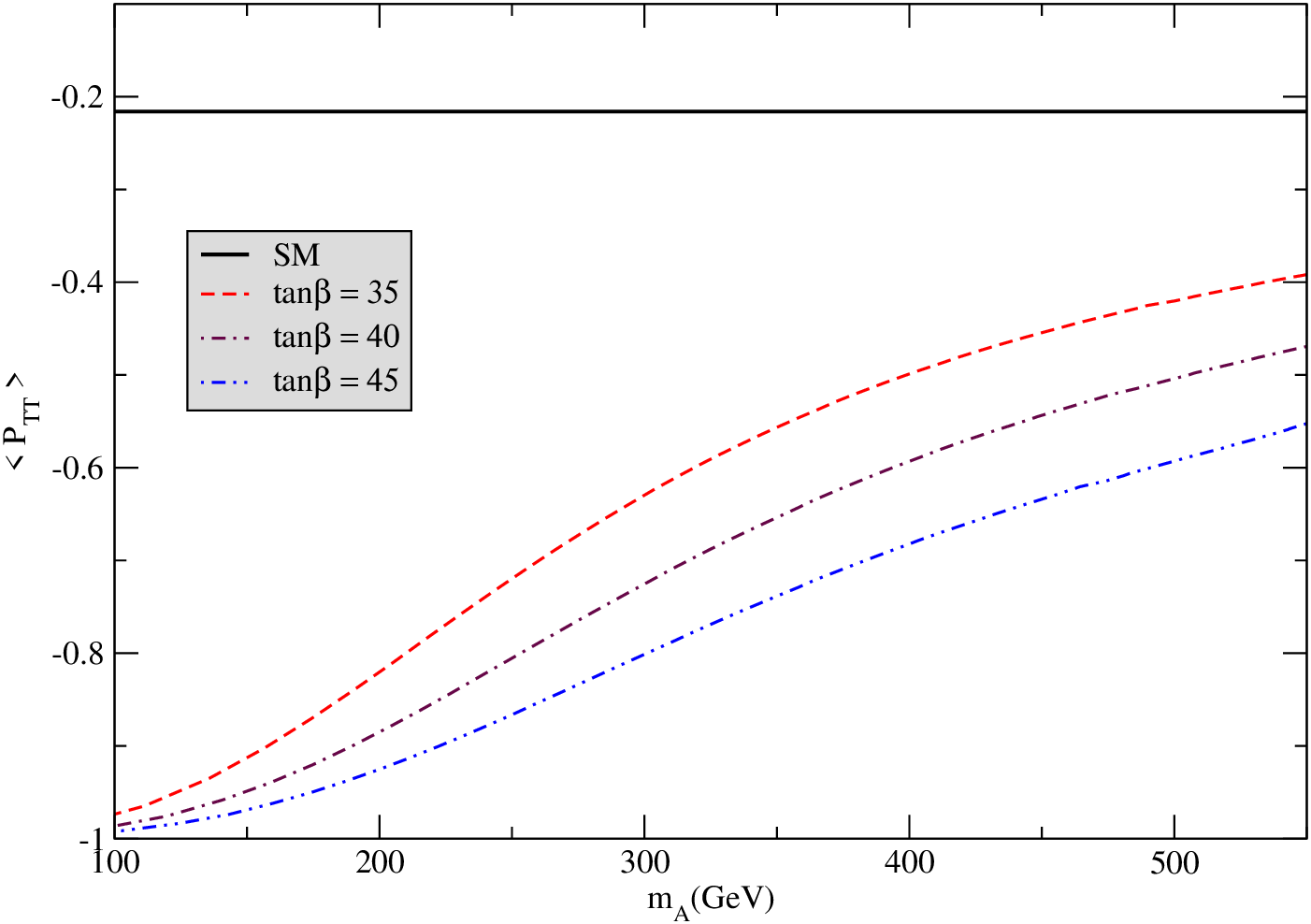}
\caption{\label{fig:ptt_ma}Integrated ${\cal P}_{TT}$ with $m_A$GeV in
rSUGRA model. Parameters same as given in Fig. \ref{fig:dr_ma}}
\end{figure*}
\end{widetext}

\end{document}